\documentclass[a4paper,11pt]{article}

\usepackage{jcappub} 
\usepackage[T1]{fontenc} 
\usepackage{array,booktabs,color,eurosym,graphicx,longtable, pbox, pdfpages, url, verbatim}
\usepackage{amsmath,amssymb,amsthm,amsxtra,overpic,bbm,bm,float,epsfig}
\usepackage{multirow}
\usepackage[figuresright]{rotating}
\usepackage{lineno}

\title{\boldmath Real-time Monitoring for the Next Core-Collapse Supernova in JUNO}

\author[6,5]{Angel Abusleme}
\author[44]{Thomas Adam}
\author[65]{Shakeel Ahmad}
\author[65]{Rizwan Ahmed}
\author[54]{Sebastiano Aiello}
\author[65]{Muhammad Akram}
\author[65]{Abid Aleem}
\author[20]{Fengpeng An}
\author[22]{Qi An}
\author[54]{Giuseppe Andronico}
% ...
\affiliation[5]{Millennium Institute for SubAtomic Physics at the High-energy Frontier (SAPHIR), ANID, Chile}
\affiliation[6]{Pontificia Universidad Cat\'{o}lica de Chile, Santiago, Chile}
% ...
\affiliation[20]{Sun Yat-Sen University, Guangzhou, China}
% ...
\affiliation[22]{University of Science and Technology of China, Hefei, China}
% ...
\affiliation[44]{IPHC, Universit\'{e} de Strasbourg, CNRS/IN2P3, F-67037 Strasbourg, France}
% ...
\affiliation[54]{INFN Catania and Dipartimento di Fisica e Astronomia dell Universit\`{a} di Catania, Catania, Italy}
% ...
\affiliation[65]{Pakistan Institute of Nuclear Science and Technology, Islamabad, Pakistan}
% ...

\vspace{+2.5cm}
\emailAdd{Juno\_pub\_comm@juno.ihep.ac.cn}

\vspace{+2.5cm}

\abstract{

The core-collapse supernova (CCSN) is considered one of the most energetic astrophysical events in the universe. The early and prompt detection of neutrinos before (pre-SN) and during the supernova (SN) burst presents a unique opportunity for multi-messenger observations of CCSN events. In this study, we describe the monitoring concept and present the sensitivity of the system to pre-SN and SN neutrinos at the Jiangmen Underground Neutrino Observatory (JUNO), a 20 kton liquid scintillator detector currently under construction in South China.
The real-time monitoring system is designed to ensure both prompt alert speed and comprehensive coverage of progenitor stars. It incorporates prompt monitors on the electronic board as well as online monitors at the data acquisition stage.
Assuming a false alert rate of 1 per year, this monitoring system exhibits sensitivity to pre-SN neutrinos up to a distance of approximately 1.6 (0.9) kiloparsecs and SN neutrinos up to about 370 (360) kiloparsecs for a progenitor mass of 30 solar masses, considering both normal and inverted mass ordering scenarios.
The pointing ability of the CCSN is evaluated by analyzing the accumulated event anisotropy of inverse beta decay interactions from pre-SN or SN neutrinos. This, along with the early alert, can play a crucial role in facilitating follow-up multi-messenger observations of the next galactic or nearby extragalactic CCSN.
}

\begin{document}
\maketitle
\flushbottom

\newpage

\section{Introduction}

The core-collapse supernova (CCSN) is considered one of the most energetic astrophysical events, accompanying the death of a massive star. A burst of neutrinos of tens of MeV energies plays important roles during its explosion and carries away most of the released gravitational binding energy of around $10^{53}$ erg. This overall picture is essentially supported by the detection of sparse neutrinos from SN 1987A in the Large Magellanic Cloud~\cite{Kamiokande-II:1987idp, Bionta:1987qt, Alekseev:1988gp}. For the next Galactic or nearby extra-galactic CCSN, more detailed time and energy spectra information of neutrinos from the CCSN are highly desired to describe and model the complex physical processes of the explosion. Such more detailed picture will be achieved by different types of modern neutrino detectors with lower energy threshold, larger target masses and complementary designs. Moreover, the first detection of neutrinos emitted prior to the core collapse (pre-SN) is also possible for nearby progenitor stars in current neutrino detectors like KamLAND~\cite{KamLAND:2015dbn} and Super-Kamiokande~\cite{Super-Kamiokande:2019xnm} as well as the future ones like Jiangmen Underground Neutrino Observatory (JUNO)~\cite{JUNO:2015zny,JUNO:2021vlw}, which would provide valuable information on the final stage of the stellar evolution and related physics probes~\cite{Kato:2020hlc,Guo:2019orq}.

{The JUNO experiment, currently being constructed in South China, is designed with a 20 kton liquid scintillator (LS) detector primarily for determining the neutrino mass ordering using reactor neutrinos~\cite{JUNO:2015zny,JUNO:2021vlw}. It is also capable of detecting both pre-supernova (pre-SN) and supernova (SN) neutrinos through the inverse beta decay (IBD) interaction, which is a dominant detection channel. Additionally, JUNO is also capable of detecting a SN burst using other channels, such as the elastic scattering processes on the electrons and free protons, as well as the charged-current and neutral current interactions on the carbon nuclei~\cite{JUNO:2015zny}. 
The detection of pre-SN and SN neutrinos provides significant opportunities to study the physics of CCSN explosions and related neutrino properties. However, to fully understand the complete picture of a CCSN event, it is essential to observe not only neutrinos but also other messengers, such as gravitational waves and electromagnetic radiation. Gravitational waves are generated during the violent and asymmetric collapse of the stellar core~\cite{Ott:2003qg}. Alongside neutrinos, they are currently the only direct real-time probes of the inner dynamics of the core collapse. On the other hand, the early electromagnetic radiation is emitted after the shock waves reach the outer envelope of the progenitor star, typically minutes to days after the neutrinos from the neutrinosphere~\cite{Adams:2013ana,Nakamura:2016kkl}. This radiation provides crucial information about the external explosion and progenitor properties. Moreover, the arrival of pre-SN neutrinos could occur several days before the CCSN explosion. 
Given that they precede the electromagnetic radiation by hours to days, the rapid detection and announcement of SN and even pre-SN neutrinos are essential for early warning for the CCSN events.
All these considerations motivate a real-time monitoring system in JUNO to provide early alerts for follow-up multi-messenger observations of the next CCSN. In addition to independent experimental efforts searching for the next CCSN, a multi-experiment collaboration, SNEWS~\cite{Antonioli:2004zb,SNEWS:2020tbu}, was created to further enhance these capabilities.}

In this work, we present the concept of the CCSN monitoring system at JUNO detecting both the pre-SN neutrinos and SN neutrinos. The system is designed with both prompt monitors on the electronic boards and online monitors at the data acquisition (DAQ) stage. These prompt monitors aim to provide much quicker alerts than the conventional online process to catch any potential multi-messenger observations. Since a Galactic or near Extragalactic CCSN is extremely rare, a specific trigger-less data processing scheme is also developed to limit the loss of CCSN-related information. We evaluate the sensitivity of the monitoring system to the CCSN, and its pointing ability using the anisotropy of the online IBD events.
Such precise directional information would be important to help the optical telescopes to catch early electromagnetic  radiation. Also, the precise onset time of signal can be useful for triangulating the source when combining information from several experiments.

The remainder of this work is organized as follows. We introduce in Sec.~\ref{sec:neutrinos} the numerical models of the Pre-SN neutrinos and SN neutrinos from progenitor stars. The JUNO detector and detection of the pre-SN and SN neutrinos are presented in Sec.~\ref{sec:juno}.
The concept of the real-time monitoring of the CCSN is depicted in Sec.~\ref{sec:Monitor}. And then the sensitivity and pointing ability of the monitoring system to the CCSN are evaluated in Sec.~\ref{sec:alert} and Sec.~\ref{sec:dir}, respectively with the simulated event samples for both pre-SN and SN neutrinos. Finally, we conclude in Sec.~\ref{sec:con}.

\section{Pre-Supernova and Supernova Neutrino Emission}
\label{sec:neutrinos}

The neutrino emission from massive stars heavier than 8 to 10 solar masses ($M_{\odot}$) commences from the fusion of hydrogen and becomes the dominant source of stellar cooling over the photon radiation following the ignition of carbon~\cite{Woosley:2002zz}. Starting from the phase of carbon burning, neutrinos are dominantly produced in pairs through the thermal processes, i.e., the plasmon decay $\gamma^* \to \nu + \overline{\nu}$, the photo-neutrino process $\gamma + e^- \to e^- + \nu + \overline{\nu}$, the pair annihilation process $e^+ + e^- \to \nu + \overline{\nu}$ and the bremsstrahlung process $e^- + Ze \to e^- + Ze + \nu + \overline{\nu}$, where $Ze$ denotes the heavy nuclei with an atomic number of $Z$.
There are also significant contributions from nuclear weak interactions after the silicon burning~\cite{Yoshida:2016imf, Patton:2017neq}, including the $e^{\pm}_{}$ capture and $\beta^{\pm}_{}$ decays of the heavy nuclei.
Prior to the core collapse, these neutrinos $\nu_{e}$, $\overline{\nu}_{e}$, $\nu_{x}$ and $\bar{\nu}_x$ ($\nu_{x}$ collectively stands for $\nu_\mu$, $\nu_\tau$ and $\bar{\nu}_x$ for their antiparticles) with energies of O(1)~MeV are usually called the pre-SN neutrinos.

At the end of the stellar evolution of massive stars, an iron core is formed and grows by the silicon shell burning. This core collapses inward under gravity once the Chandrasekhar mass limit is reached, which results in the subsequent CCSN~\cite{Mirizzi:2015eza}. A neutrino burst (called SN neutrinos) of $\sim$10~s with average energy of tens of MeV and higher luminosity than pre-SN neutrinos are expected to be produced in three main phases~\cite{Janka:2017vlw,Burrows:2020qrp,Nakazato:2012qf}, namely, the shock breakout burst phase, the post-bounce accretion phase and the proto-neutron star cooling phase.
The SN neutrinos are neither obscured by the interstellar dust as the electromagnetic radiations, nor totally absent for failed explosions resulting in a black hole, highlighting the importance of neutrino detection for the observation of the CCSN.

In this work, we employ different numerical models for the fluxes of pre-SN neutrinos and SN neutrinos to study the influence of different models. The pre-SN models are from Patton \textit{et al.}~\cite{Patton:2017neq} for the $15\ M_{\odot}$ and $30\ M_{\odot}$ progenitor stars, where both thermal processes and nuclear weak interactions are taken into account in the pre-SN simulation. The SN neutrino models are provided by the Nakazato group~\cite{Nakazato:2012qf} and the Garching group~\cite{Hudepohl:2013zsj}. The Nakazato models are simulated for progenitor masses of $13\ M_{\odot}$ and $30\ M_{\odot}$ with metallicities and shock revival times of (0.004, 100ms) and (0.002, 300ms) respectively. The Garching models are simulated up to the cooling phase with the equation of state LS220, including the convection via a mixing-length scheme and the effect of nucleon potentials on the neutrino opacity for the progenitor masses of $11.2\ M_{\odot}$ and $27\ M_{\odot}$. Note that the time duration of Nakazato models lasts for about 20 s while that of Garching models cuts off at about 3 to 4~s.
All the pre-SN and SN models provide the time-dependent luminosity and energy spectra for different flavor neutrinos. 
The neutrino fluxes at the detector can be obtained with the scale factor of $1/D^{2}$ with $D$ being the distance to the progenitor star of CCSN.
It should be noted that the definitions for the zero point of time are different in the pre-SN and SN models, where it refers to the core collapse time for pre-SN models while it is defined as the shock bounce time for SN models.

The pre-SN neutrinos and SN burst neutrinos undergo a flavor conversion when propagating from the core to the terrestrial detectors. In the pre-SN phase, collective oscillations~\cite{Duan:2010bg,Chakraborty:2016yeg} can be neglected due to the low number density of neutrinos in the medium, but that is not the case for SN neutrinos, due to the large quantity of neutrinos produced during the explosion. From the neutrino-sphere to around one thousand kilometers, the neutrino-neutrino refraction may lead to the energy spectral split of SN neutrinos~\cite{Duan:2010bg}. However, it is still unclear whether the collective neutrino oscillations do happen in a real supernova environment~\cite{Chakraborty:2016yeg}. For simplicity, we temporarily neglect the collective neutrino oscillations, but one should keep in mind that they may have important effects on the detection of SN neutrinos.
Even farther out from the neutrino-sphere, the Mikheyev-Smirnov-Wolfenstein (MSW) matter effects~\cite{Wolfenstein:1977ue,Mikheyev:1985zog} will play an important role and leave an imprint on the neutrino spectra~\cite{Dighe:1999bi}.
According to the current neutrino oscillation data~\cite{ParticleDataGroup:2018ovx}, the resonant flavor conversions for both pre-SN and SN neutrinos are highly adiabatic and depend on the neutrino mass ordering. Therefore the neutrinos fluxes $F^{}_{\nu}$ at the Earth can be expressed in terms of the initial ones $F^0_{\nu}$ at production in the following forms~\cite{Dighe:1999bi}:
\begin{eqnarray}
F^{}_{\nu^{}_e} &=& p F^0_{\nu^{}_e} + (1 - p) F^0_{\nu^{}_x}\; ,\\
F^{}_{\overline{\nu}^{}_e} &=& \bar{p} F^0_{\overline{\nu}^{}_e} + (1 - \bar{p}) F^0_{\overline{\nu}^{}_x}\; ,\\
F^{}_{\nu^{}_x} &=& 0.5 (1 - p) F^0_{\nu^{}_e} + 0.5 (1 + p) F^0_{\nu^{}_x}\; ,\\
F^{}_{\overline{\nu}^{}_x} &=& 0.5 (1 - \bar{p}) F^0_{\overline{\nu}^{}_e} + 0.5 (1 + \bar{p}) F^0_{\overline{\nu}^{}_x}\; ,
\label{eq:osc}
\end{eqnarray}
where $p = \sin^2 \theta^{}_{13} \approx 0.022$ and $\overline{p} = \cos^2 \theta^{}_{13} \cos^2 \theta^{}_{12} \approx 0.687$ for the normal neutrino mass ordering (NO), and $p = \sin^2\theta^{}_{12} \cos^2\theta^{}_{13} \approx 0.291$ and $\overline{p} = \sin^2 \theta^{}_{13} \approx 0.022$ for the inverted neutrino mass ordering (IO). 
The upper panels of Fig.~\ref{fig:espec} show the examples of neutrinos fluxes for pre-SN at 0.2~kpc and SN at 10~kpc respectively. Finally, Earth matter effects are not included for either the pre-SN neutrinos or SN neutrinos.

\section{JUNO Detector and Neutrino Signals}
\label{sec:juno}

JUNO is a multi-purpose neutrino experiment under construction at Jiangmen in Guangdong Province, China.
The complete detector of JUNO consists of the central detector (CD), the water pool (WP) and the top tracker (TT)~\cite{JUNO:2023cbw}, all are placed about 700 m underground for shielding from cosmic rays.

The CD is composed of 20 kt LS contained in a spherical acrylic vessel with a 35.4 m inner diameter. The LS is composed of linear alkylbenzene as the solvent, 2,5-diphenyloxazole (PPO) as the fluor and p-bis-(o-methylstyryl)-benzene (bis-MSB) as the wavelength shifter~\cite{JUNO:2020bcl}. It is the main target for the pre-SN and SN neutrinos at JUNO. To detect the light induced by the final-state particles of neutrino interactions, 17,612 20-inch PMTs (LPMTs) and 25,600 3-inch PMTs are installed on the outer stainless steel latticed shell of CD with 40.1~m diameter. JUNO can reach an excellent energy resolution of about 3\%@1 MeV~\cite{JUNO:2020xtj} and its vertex resolution can be 10 cm at 1 MeV~\cite{Qian:2021vnh}. The veto system with the WP and TT is designed to tag the muons with high efficiency and suppress the cosmogenic backgrounds, where the WP contains 35 kton ultrapure water and serves also as a passive shielding for the radioactivity from surrounding rocks.
Its muon detection efficiency is greater than 99\% with about 2,400 20-inch PMTs installed on the outer surface of the stainless steel Shell Structure of the CD. The TT reuses the plastic scintillating strips from the OPERA experiment~\cite{Adam:2007ex} and is placed on top of the WP detector to track muons.

JUNO can achieve a low energy threshold for the purpose of the pre-SN and SN neutrino detection. Currently the default global trigger strategy of CD applies a typical multiplicity trigger algorithm. It can efficiently suppress the backgrounds mostly from the dark noise and intrinsic radioactivity of $^{14}{\rm C}$ and can reach an energy threshold of ${\rm O}(0.1)$~MeV~\cite{Fang:2019lej}. 

\begin{figure}
\begin{center}
\begin{tabular}{l}
\includegraphics[width=0.51\textwidth]{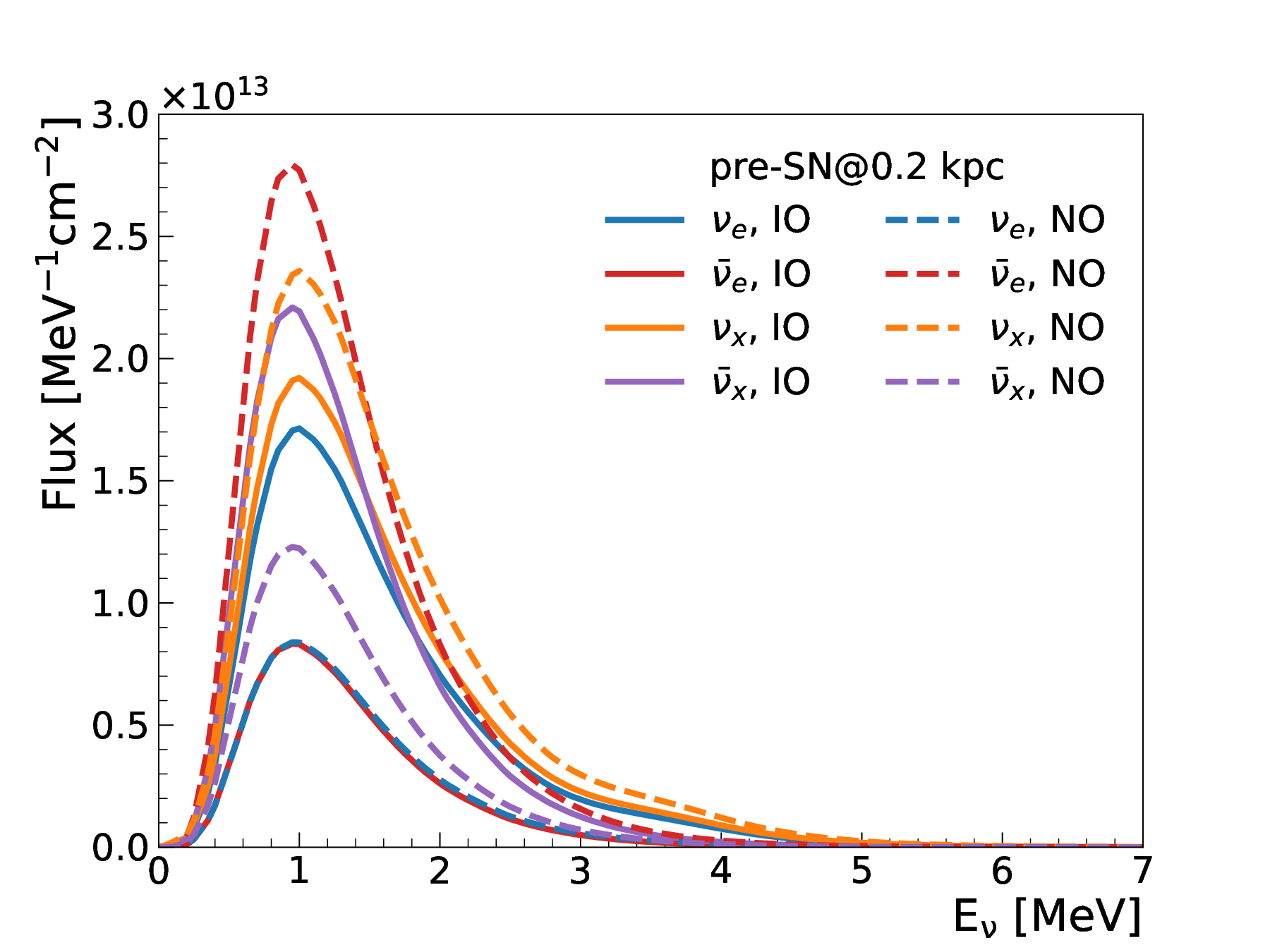}
\includegraphics[width=0.51\textwidth]{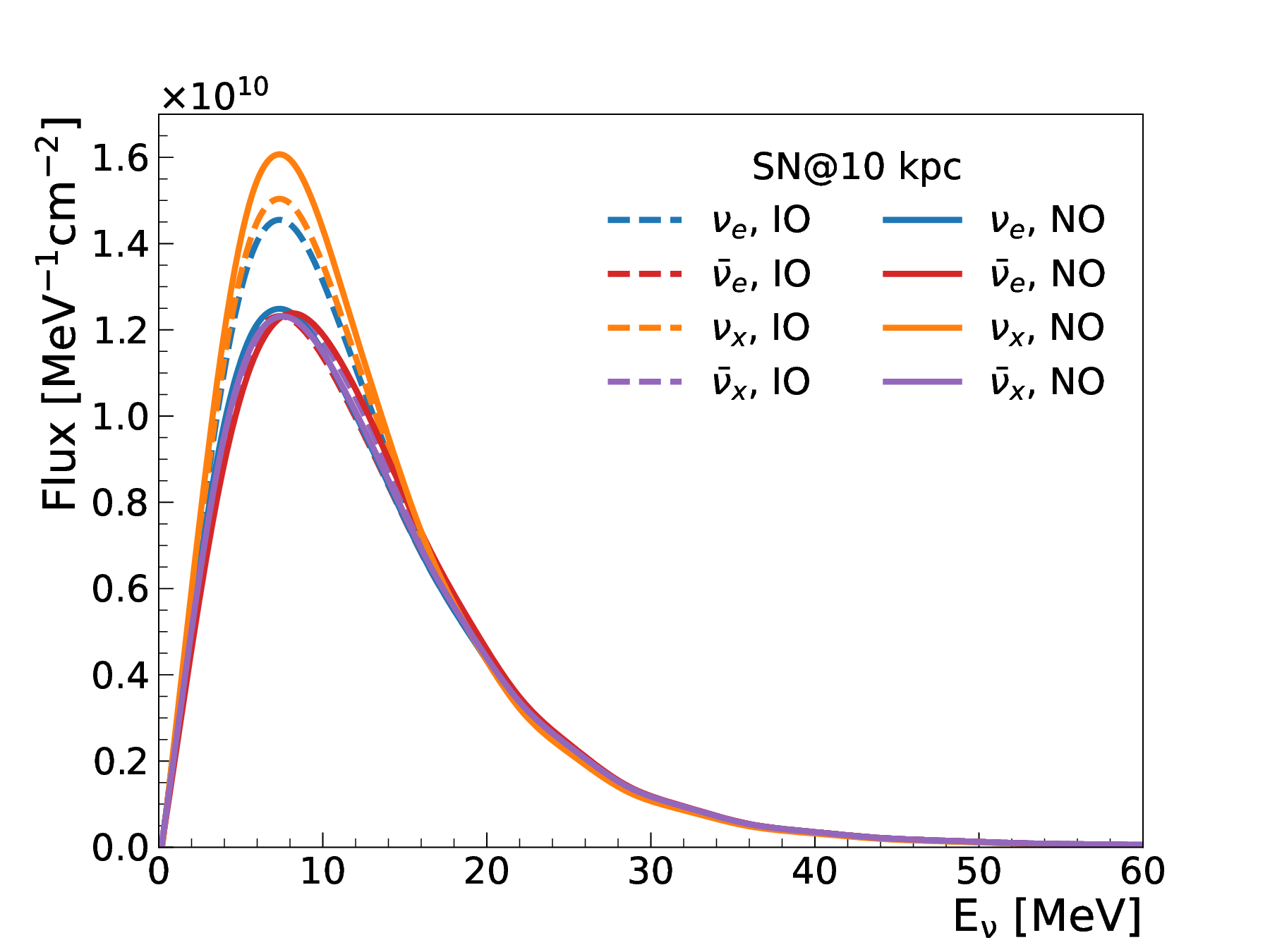} 
\\
\includegraphics[width=0.51\textwidth]{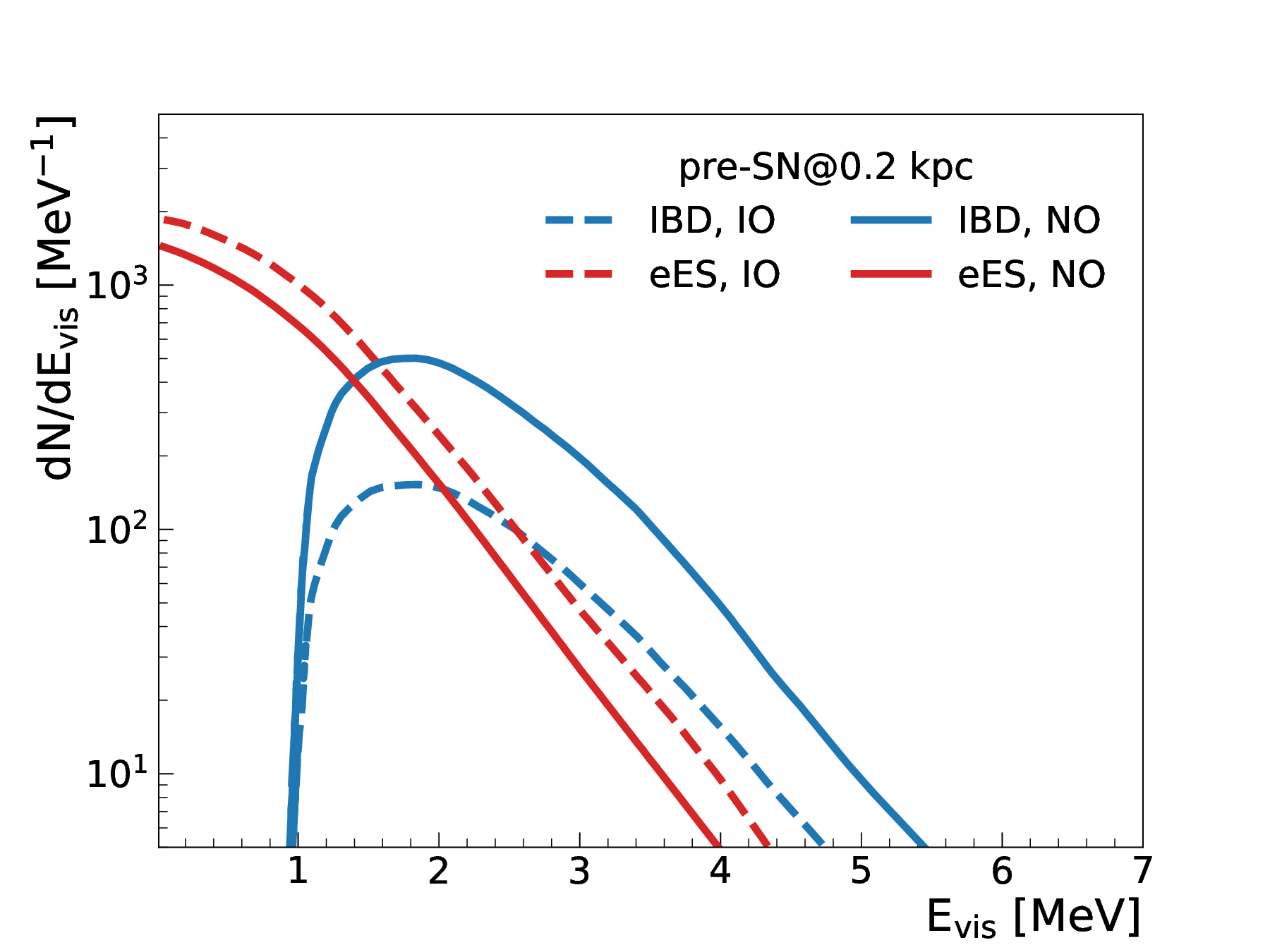}
\includegraphics[width=0.51\textwidth]{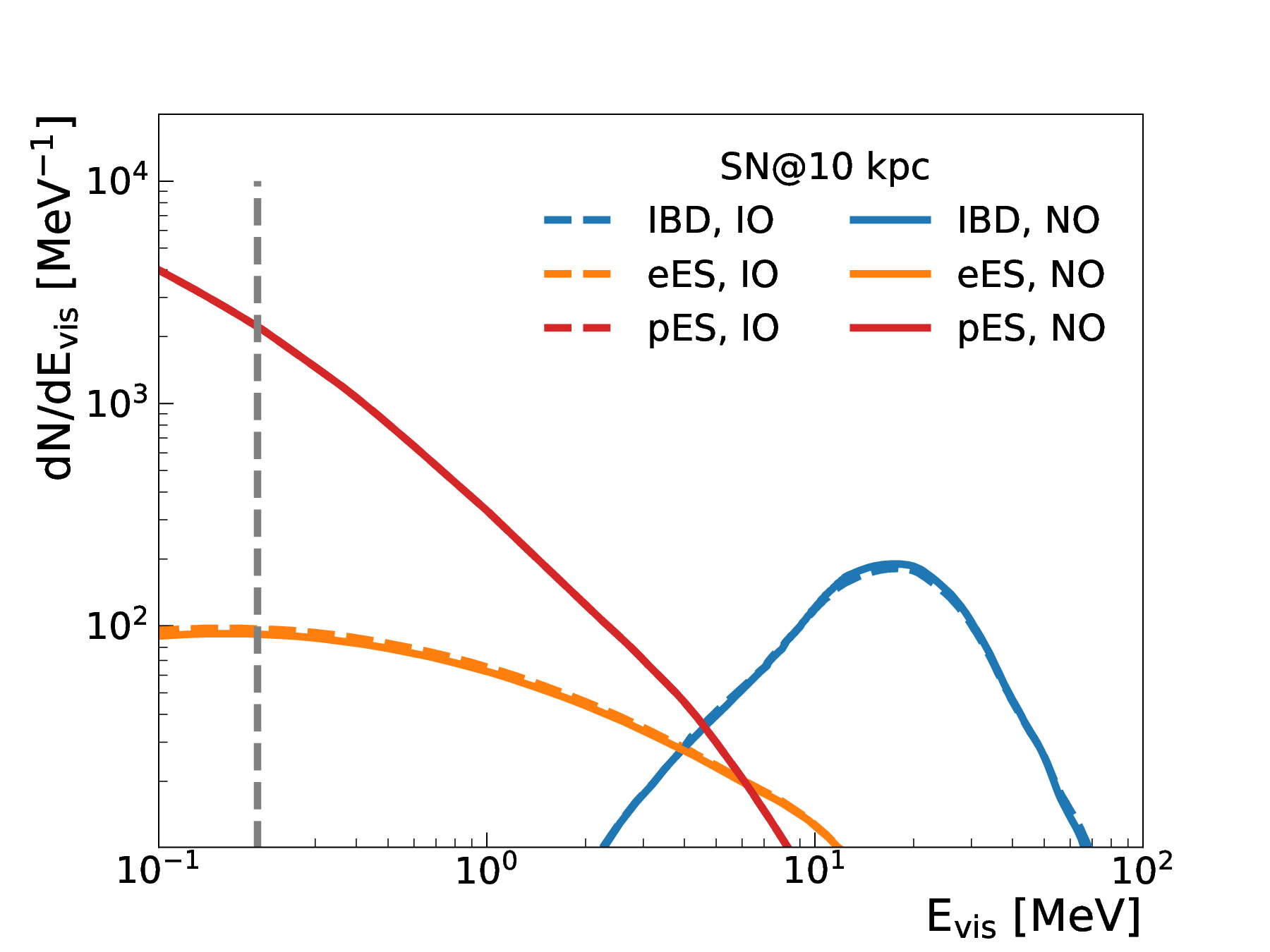}
\end{tabular}
\end{center}
\vspace{-0.6cm}
\caption{The neutrino fluxes (upper panels) and the expected visible energy spectra (bottom panels) for pre-SN neutrinos of 30 $M_{\odot}$ Patton model at 0.2~kpc integrated over the last 5 days before the core collapse (left panels) and for SN neutrinos of 30 $M_{\odot}$ Nakazato model at 10 kpc (right panels), estimated with 20~kton LS for both the NO and IO cases. The IBD prompt energy from the annihilated positron is related to the neutrino energy as $E_{\rm vis}\approx E_{\nu}-0.8$~MeV. For the $e$ES channel, $E_{\rm vis}$ is estimated with the kinetic energy of the recoiled electron. For the $p$ES channel, $E_{\rm vis}$ refers to the light emitted from the quenched recoiled proton. The corresponding detector response between $E_{\rm vis}$ and $E_\nu$ can be found in~\cite{Li:2019qxi}
\label{fig:espec}}
\end{figure}

The energies of SN neutrinos are several tens of MeV, so in JUNO, all flavors of SN neutrinos can be registered via multiple channels. The dominant channel, which exists exclusively for $\bar{\nu}_e$, is the IBD process $\overline{\nu}_{e}+p \rightarrow e^{+}+n$, with a coincidence signature of the prompt signal from the positron's kinetic energy deposition and annihilation and the delayed signal from 2.2~MeV $\gamma$ from the neutron capture on hydrogen. The subdominant channels are neutrino-proton elastic scattering ($p$ES), $\nu+p \rightarrow \nu+p$, and neutrino-electron elastic scattering ($e$ES), $\nu+e^{-} \rightarrow \nu+e^{-}$, with single visible signals from energy deposition of the proton or electron. In addition, the charged-current (CC) and neutral-current (NC) interactions of neutrinos on $^{12}{\rm C}$ nuclei are also accessible for SN neutrinos of the higher energy range. The energies of pre-SN neutrinos are in the order of O(1)~MeV. So the interactions of neutrinos on the carbon nuclei are not accessible for pre-SN neutrinos. Meanwhile, the visible energy of $p$ES events are below threshold due to the severe proton quenching effect. Thus only the IBD and $e$ES interactions are available for pre-SN neutrinos.

\begin{figure}
\begin{center}
\begin{tabular}{l}
\hspace{-1cm}
\includegraphics[width=0.53\textwidth]{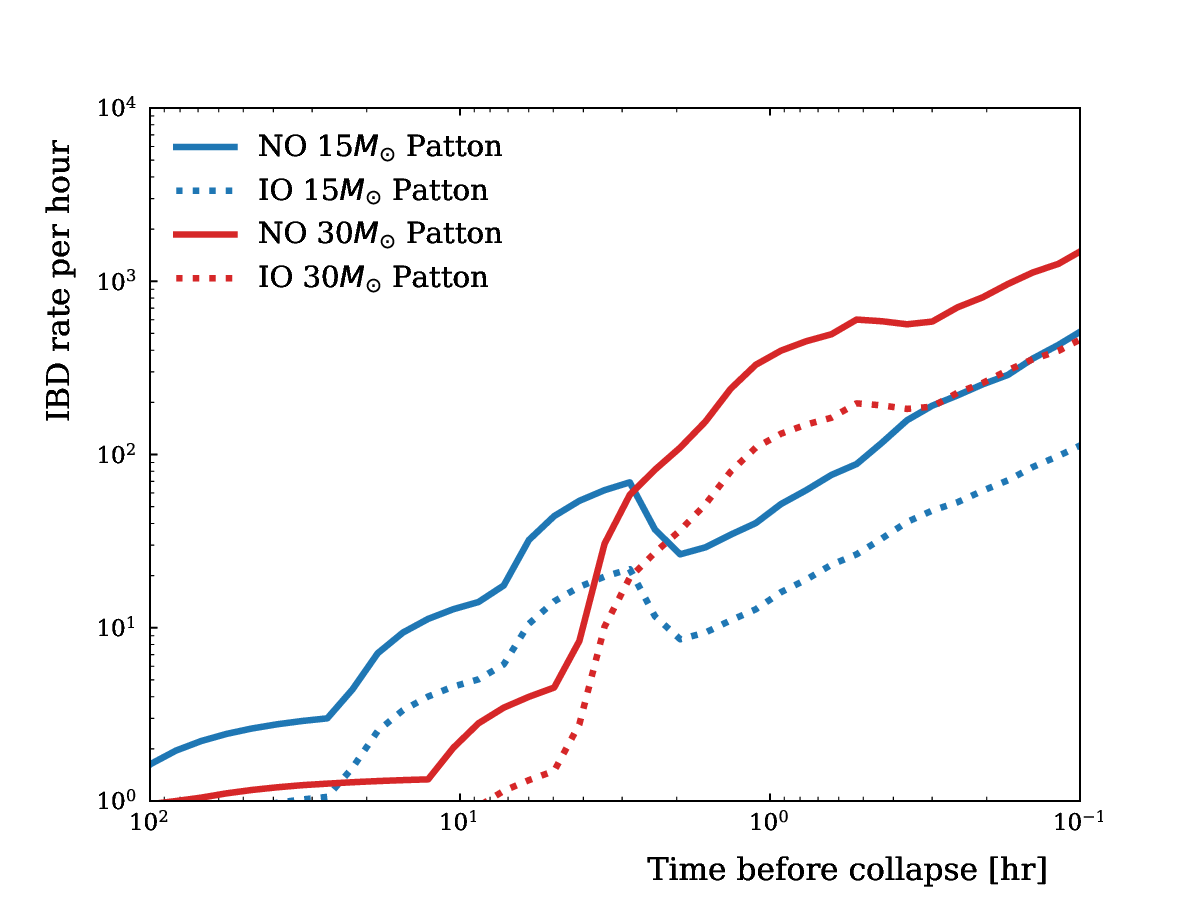}
\includegraphics[width=0.53\textwidth]{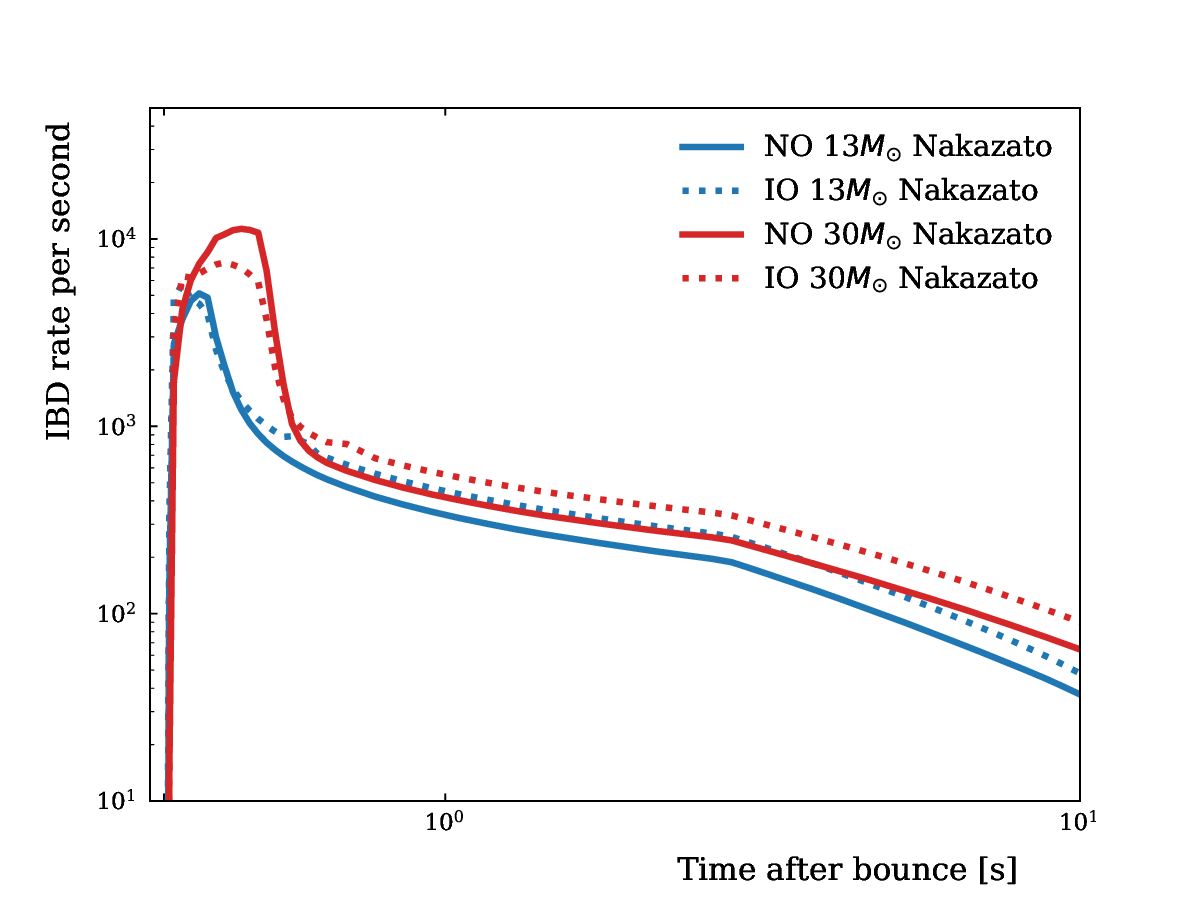}
\end{tabular}
\end{center}
\caption{The IBD time evolution of pre-SN neutrinos with Patton models for a progenitor at 0.2 kpc before collapse (left) and SN neutrinos with Nakazato models for a CCSN at 10 kpc (right). Here, 0.2 MeV energy threshold is assumed. The same definition of IBD events is applied as in Fig.~\ref{fig:espec}.}
\label{fig:IBDt}
\end{figure}

The neutrino fluxes of different pre-SN and SN models with the MSW effect considered are convoluted with the energy-dependent interaction cross sections for the IBD~\cite{Strumia:2003zx}, $p$ES~\cite{Ahrens:1986xe}, $e$ES~\cite{Radel:1993sw} and the neutrino interactions on the carbon nuclei~\cite{Fukugita:1988hg}. 
The final-state particles of the pre-SN and SN interactions are generated in the time sequence and propagated to the JUNO detector simulation. In Fig.~\ref{fig:espec} we illustrate the expected visible energy spectra in the JUNO LS detector for the IBD and $e$ES interaction channels of the pre-SN neutrinos using the $30 M_{\odot}$ Patton model integrated over the last 5 days before core collapse at 0.2 kpc (left panel), and for the dominant IBD, $e$ES and $p$ES interaction channels of the SN neutrinos using the $30 M_{\odot}$ Nakazato model at 10 kpc (right panel) for both NO and IO cases. With a 0.2 MeV energy threshold assumed, the time evolution of the expected IBD events for pre-SN neutrinos (left panel) and SN neutrinos (right panel) are illustrated in Fig.~\ref{fig:IBDt}. 
The event rate increases gradually within several days before collapse and undergoes a sudden increase within a few seconds as core collapse. This characteristic can be used to distinguish whether a CCSN happens.

The IBD is a golden channel for both pre-SN neutrinos and SN burst neutrinos, thanks to a large number of free protons in the LS medium and largest cross section at MeV energies. The IBD events can be tagged with a high efficiency because of the coincident prompt and delayed signals.
For the pre-SN neutrinos, the number of IBD events depends crucially on the neutrino mass ordering, which is suppressed in the IO case by a factor of three or four compared with that in the NO case.
The $p$ES and $e$ES elastic scattering channels are sensitive to all flavor neutrinos with smaller cross sections than that of the IBD channel.
For the SN neutrinos, they dominate in the lower energy range below the IBD threshold, which implies that the energy threshold of the LS detector plays an important role for the extraction of $p$ES and $e$ES events. Although the event separation of $p$ES and $e$ES events are more difficult than the IBD channel because they are single signals, it still can be realized by using the pulse shape discrimination method~\cite{Rebber:2020xfi}.
For the pre-SN neutrinos, most of the $e$ES events appear in the energy range below 2~MeV, which are highly contaminated with the backgrounds from radioactivity and cosmogenic isotopes. 
In contrast to IBD events, the numbers of $e$ES events are much higher in the IO case than that in the NO case, which provides an opportunity to probe the mass ordering with pre-SN neutrinos~\cite{Guo:2019orq}. 
In this paper, IBD events are mainly considered in the monitoring systems, while other channels will be considered in the future.

\section{Concept of the CCSN Monitoring System}
\label{sec:Monitor}

The real-time monitoring system of the CCSN in JUNO aims to provide early alerts and record the CCSN-related data as much as possible for the next Galactic or near extragalactic CCSN. It features a redundant design consisting of the prompt monitor embedded in both the global trigger and Multi-messenger trigger electronic boards, and the online monitor at the DAQ stage.
The prompt monitor aims to reduce the timing latency of issuing an alert for a CCSN event, while the online monitor operates with the finely reconstructed information at the DAQ stage and can extract the fundamental CCSN characteristics in a rapid and effective way.

The pre-SN neutrinos and SN neutrinos will produce an increase of the event rate compared to the rate of the continuous stable sources, e.g., reactor neutrinos, solar neutrinos, radioactivity and cosmic muons. By monitoring the rate of events with characteristics similar to those expected from a CCSN, this real-time monitoring system will release an alert once a change of the steady event rate is identified. More details will be discussed in Sec.~\ref{sec:alert}. 
Then an early alert will be sent to the internal collaboration and astronomical communities for follow-up multi-messenger observations of the CCSN.
In the remaining parts of this section, more detailed description of the system will be introduced, including the PMTs readout and trigger electronics, the prompt monitor and online monitor systems.

\subsection{Large PMT readout electronics}

As discused in~\cite{JUNO:2021vlw} the analog signal coming from the JUNO 20-inch PMTs is fed to a readout electronics board, the so-called Global Control Unit (GCU) which amplifies the PMT analog signal and converts it to a digital waveform with a 14 bit ADC and a sampling rate of 1 GS/s. Each GCU handles three 20-inch PMTs, in parallel, and it is installed on the CD support structure, very close to the PMTs, inside a water-tight box. The digitized PMT waveform is further processed in a local FPGA and trigger primitives are generated. All local trigger information are send to the electronics rooms where they are validated by the Central Trgger Unit (CUT) and, in parallel elaborated by the Multi-messenger trigger boards. The CTU validates the trigger signal (based on simple multiplicity requirements or more complicated event topology) and issue the trigger validation for the waveform to be read-out. In parallel to the single PMT waveform read-out stream, the waveform integrated charge is computed on each GCU FPGAs and sent with the event timestamp, in asynchronous mode, to the DAQ for each PMT signal. The latter stream allows a higher data rate due to the reduced size of the transmitted information. Finally, in order to handle exceptional high rate events, like those coming from a nearby Galactic CCSN, a dedicated 2 GBytes DDR memory has been placed on each GCU readout board to provide a local storage of the events in case of a nearby Galactic CCSN. According to evaluation and readout tests, the readout system will be able to handle and record all the events for a CCSN explosion at 0.5 kpc or further.

\subsection{Prompt Monitor on Trigger Boards}
\label{sec:PMgt}

The prompt monitoring system runs on FPGAs of the electronic boards in the global trigger system, which is aimed to reduce the time latency of issuing an alert for the followup multi-messenger observations of the next CCSN. It can continuously monitor core-collapse supernovae by detecting an increase in event rate of supernova neutrinos.

The information of fired LPMTs collected from GCUs is processed on the global-trigger board with a 16 ns clock cycle.
An \textit{event} is identified based on the PMT multiplicity condition or a more refined topological distribution of the fired LPMTs~\cite{Fang:2019lej}. The acquisition is anticipated to have an energy threshold of $O(0.1)$ MeV, where the backgrounds from the radioactivity of $^{14}{\rm C}$ and coincidence of LPMT dark noises are mostly filtered. In normal operation, the global-triggered event rate is approximately 1~kHz in CD.

Once there is a global-triggered event in CD, the total number of LPMT hits of this event $N_{\rm hit}$ is counted within a time window of $T_{\rm hit}$ (e.g., 1 $\mu$s) opened right after the global-triggered timestamp. Obviously, $N_{\rm hit}$ is generally proportional to the visible energy of the particle deposited in LS, which can be used to discriminate SN neutrinos with energies of $O(10)$~MeV from other kinds of neutrino sources and the radioactivity signals. Furthermore, in order to suppress background induced by cosmic muons, the trigger information of WP is introduced to the global-trigger board in real time. The WP trigger is activated by a multiplicity algorithm with a threshold value larger than the number of fired LPMTs caused by SN neutrinos in WP. The global-triggered CD events are vetoed for a time period of 1.5~ms after the WP trigger timestamp.
With SN neutrinos as the diagnostic tool of the CCSN, monitoring candidates are selected outside the veto period with $N_{\rm low}<N_{\rm hit}<N_{\rm high}$, where the current $N_{\rm high}$ and $N_{\rm low}$ are chosen to map onto the total numbers of hits corresponding to the visible energies of around 10~MeV and 40~MeV respectively. In such a scenario, events in proximity of the calibration source are excluded from the prompt monitor during regular calibration~\cite{JUNO:2020xtj}.
When a critical increase in the event rate is detected in the time series of SN neutrino candidates, it will trigger a prompt alert of the candidate CCSN event and inform the DAQ server and calibration system, stopping the movement of calibration sources and recording the position and status of them for future data analysis. Immediately the onsite computer, the DAQ manager and calibration system will be informed, and then the DAQ servers will store the corresponding data that will be described in Sec.~\ref{sec:DAQ}.

\begin{figure}
\begin{center}
\begin{tabular}{c}
\hspace{-1cm}
\includegraphics[scale=0.24]{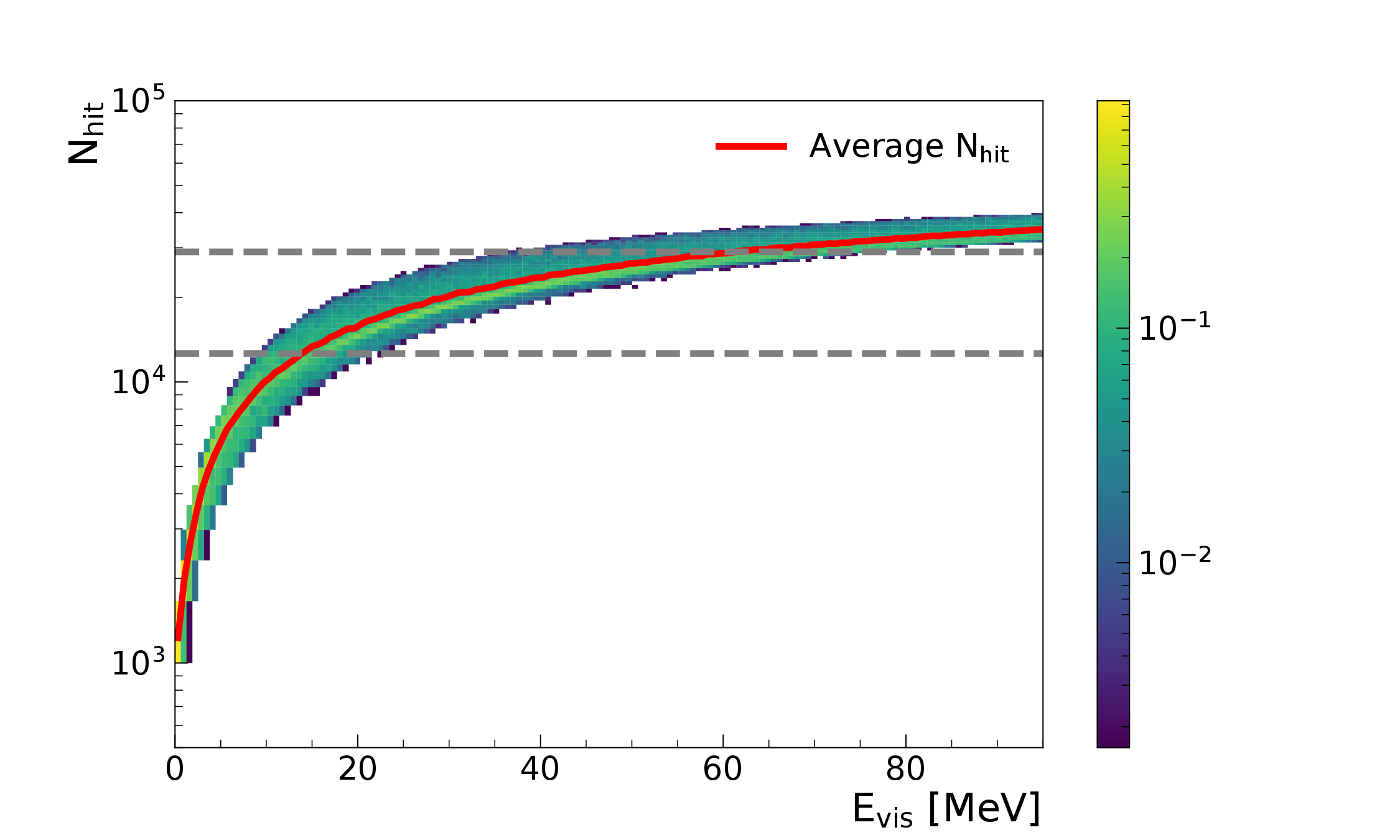}
\hspace{-1.45cm}
\includegraphics[scale=0.24]{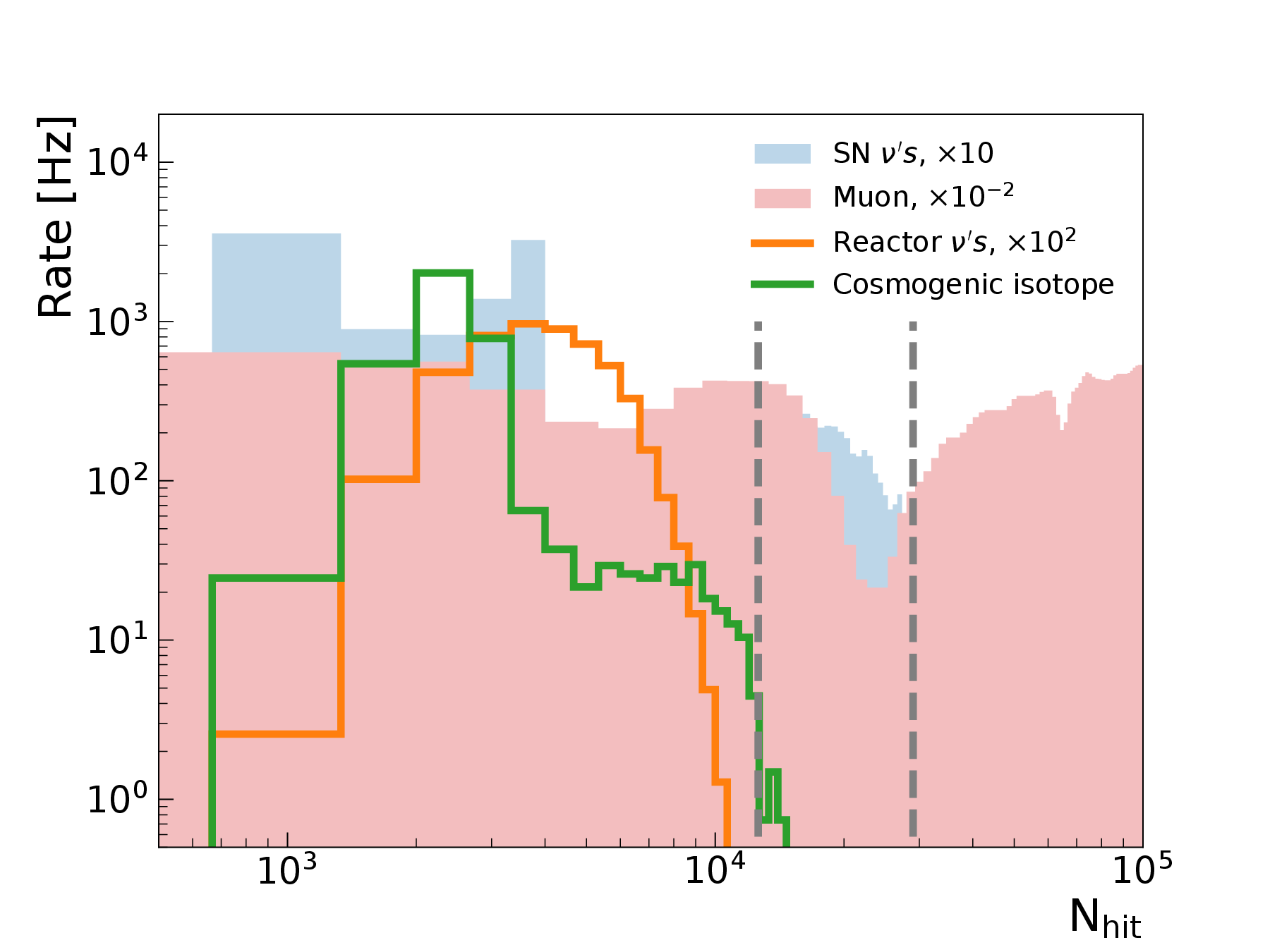}
\end{tabular}
\end{center}
\vspace{-0.5cm}
\caption{(Left panel) The number of LPMT hits $N_{\rm hit}$ versus the visible energy of positrons.
The red line is the average number of LPMT hits with respect to the visible energy,
The color bar represents the probability of positrons with the energy $E_{\rm vis}$
to produce a specific number of hits, 
$N_{\rm hit}$, in the LPMTs. (Right panel) The $N_{\rm hit}$ distributions from different triggered sources with no muon veto: the prompt signals of reactor neutrinos, muons going through CD, cosmogenic isotopes and SN neutrinos using Nakazato model with $30~M_{\odot}$. The dashed lines indicate the selection criteria of prompt monitor candidates corresponding to false alert rate being 1/month. Note that the triggered rate of SN neutrinos has been averaged over the CCSN duration of 10 s in the right panel.}
\label{fig:nPMTsPM}
\end{figure}
The relationship between $N_{\rm hit}$ and visible energy $E_{\rm vis}$ in LS is evaluated with the simulated positron samples, as is shown in the left plot of Fig.~\ref{fig:nPMTsPM}. The two dashed lines correspond to $N_{\rm low}$ and $N_{\rm high}$, which are used to select SN neutrino candidates. Tab.~\ref{tab:summary} summarizes the expected numbers of the signal candidates induced by SN neutrinos using different numerical models, which dominantly come from the prompt signals of IBD interactions (about 80\%) and partly from the $e$ES and neutrino interactions on $^{12}{\rm C}$ (about 20\%). Also, the right plot of Fig.~\ref{fig:nPMTsPM} illustrates the distribution of $N_{\rm hit}$ from different sources, including all signals of SN neutrinos, the prompt signal of reactor neutrinos, muons going through CD and long life-time cosmogenic isotopes from Ref.~\cite{JUNO:2020hqc}. Here, the event rates are scaled by different factors for comparison and that for SN is averaged over the CCSN duration of 10~s. There are many other signals which have either small energy or negligible event numbers in the time duration of $O(10)$~s, so that their contributions to the background can be neglected. {Note that muon veto is not applied here, and thus event rate from the muon induced background is much higher than that of the SN neutrino candidate. Hence a 1.5~ms muon veto is introduced to significantly decrease this component.}
The muon rate at the JUNO detector is about 3.9~Hz with an average energy of 207~GeV~\cite{JUNO:2015zny}. The WP trigger designed for the prompt monitor can tag the muons passing through the WP detector with the track length greater than 1 m with an efficiency larger than 99.5\%. The un-tagged muons are mainly due to the corner clipper muons with a short track length and the muons only passing through the surrounding rocks. Fast neutrons produced by these un-tagged muons in rocks may reach the LS detector and induce global-triggered events. They are evaluated with detailed simulations in the JUNO framework and can introduce a background rate of less than 13 per day in the prompt monitor.
For the tagged muons, the background candidates after a 1.5~ms veto are mainly from long life-time isotopes of the cosmic muon spallation process. In order to satisfy the false alert rate (see Sec.~\ref{sec:alert}) requirements of the prompt monitoring system, two different values of $N_{\rm low}$ are chosen to suppress background. Hence, referring to the isotope yields in Ref.~\cite{JUNO:2020hqc}, the expected background rate is estimated to be about 83/day or 45/day according to the $N_{\rm low}$ cuts, with contributions from the $^{12}$B, $^8$B, $^8$Li, $^9$C, $^9$Li, $^{11}$Be nuclei and the fast neutron.

\subsection{Online Monitor at the DAQ Stage} 
\label{sec:DAQ}

In parallel to the prompt monitoring system on the electronic boards, another online monitor system will be developed at the DAQ stage to maximize the alert coverage of the progenitor stars. The online monitor is implemented at the software level by utilizing the finely reconstructed events. It will operate in full time and support two online monitors for either the pre-SN neutrinos or SN neutrinos to diagnose the candidate CCSN event. A specific data flow and the triggerless data-saving strategy are included to maximize recording of CCSN-related data. 
Below, we can see that the online monitor performs event reconstruction and selects IBD events, which can greatly reduce backgrounds, especially for low-energy pre-SN neutrinos. Hence, the online monitor can also utilize pre-SN neutrinos for early warning, while the prompt monitor can not due to the lack of reconstructed information.

Let us first introduce the design of the data processing flow. The data streams from all sub-detectors are collected by the DAQ servers. The triggerless data of LPMTs in CD are grouped into events based on the online software trigger, and further packed with data from other sub-detectors for online event reconstruction. The software trigger algorithm for the triggerless data of CD can achieve an energy threshold of $O(0.1)$ MeV, which is similar to that of the global trigger strategy, but it can be further optimized in the future since all the raw LPMT hit information is available. To reduce the data size, the online fast filtering algorithm is first used and then the delicate reconstruction to get the event information, such as the energy and vertex of a CD event or the muon track in the veto system. These reconstructed events are assembled in time sequence and used in the online \textit{Pre-SN Monitor} and \textit{SN Monitor}.

As discussed in Sec.~\ref{sec:juno}, the pre-SN and SN neutrinos are detectable in LS via several distinct neutrino interaction channels. A sudden increase of event rates from these channels can help to diagnose a CCSN candidate. In this work, we concentrate on the online reconstructed information of IBD events to evaluate the baseline performance for both the \textit{Pre-SN Monitor} and \textit{SN Monitor}. Further optimization can be obtained with more interaction channels. {The signals generated by the IBD process exhibit distinct energy, temporal and spatial characteristics. The energy of the fast signal reflects the energy of the neutrino, while the energy of the delayed signal originates from the 2.2 MeV gamma released by neutron capture. The characteristic capture time of the neutron is approximately {200 \textmu{}s}, and it typically travels a distance of less than 2 m in the liquid scintillator before being captured. By utilizing these features, IBD signals can be efficiently selected.} The IBD event candidates for the \textit{Pre-SN Monitor} (denoted as "$pre$IBD") are selected in CD with {the following} selection criteria: 
\begin{itemize}
\item Event vertex must be in the fiducial volume with $r < 17.2$ m,
\item The prompt signal is required to be within the energy window of 0.7 MeV $< E_{\rm p} < $3.4~MeV and the delayed signal with 1.9 MeV $< E_{\rm d} <$ 2.5~MeV.
\item The coincidence of prompt and delayed signals are further filtered with their time interval $\Delta T_{\rm d-p} < $1.0 ms and the vertex distance $R_{\rm d-p} < 1.5$ m. 
\end{itemize}
And the same muon veto criteria as in Ref.~\cite{JUNO:2022mxj} are used:
\begin{itemize}
\item For muons tagged by the WP detector, veto the whole LS volume for 1.5 ms.
\item For muons with well reconstructed tracks in CD, veto those candidate events with reconstructed vertices less than 1 m, 2 m, and 4 m away from the corresponding muon tracks for 0.6 s, 0.4 s and 0.1 s, respectively. For muons without properly reconstructed tracks, veto the whole LS volume for 0.5 s.
\item A 1.2 s veto is applied for all candidate events reconstructed inside a sphere with the radius of 3 m around the reconstructed vertex for spallation neutron captures.
\end{itemize}
The IBD event candidates for the \textit{SN Monitor} (denoted as "$sn$IBD") are selected with similar IBD criteria as $pre$IBD except for a different prompt energy cut of $E_{\rm p}>E_{\rm low}$, where $E_{\rm low}$ uses different values to satisfy the false alert rate requirement (5.5~MeV and 7.5~MeV for 1/month and 1/year) like the prompt monitor case.
The muon veto criteria for $pre$IBD is not feasible to $sn$IBD any more due to the short burst time of $O(10)$ s. The muon veto criteria for the $sn$IBD selection is only to veto the whole LS volume for 1.5 ms if muons are tagged by the WP detector. During the calibration period, these candidates will be further selected outside of the default calibration regions, which is feasible given the effective communication between the calibration and DAQ systems.

\begin{figure}
\begin{center}
\includegraphics[scale=0.5]{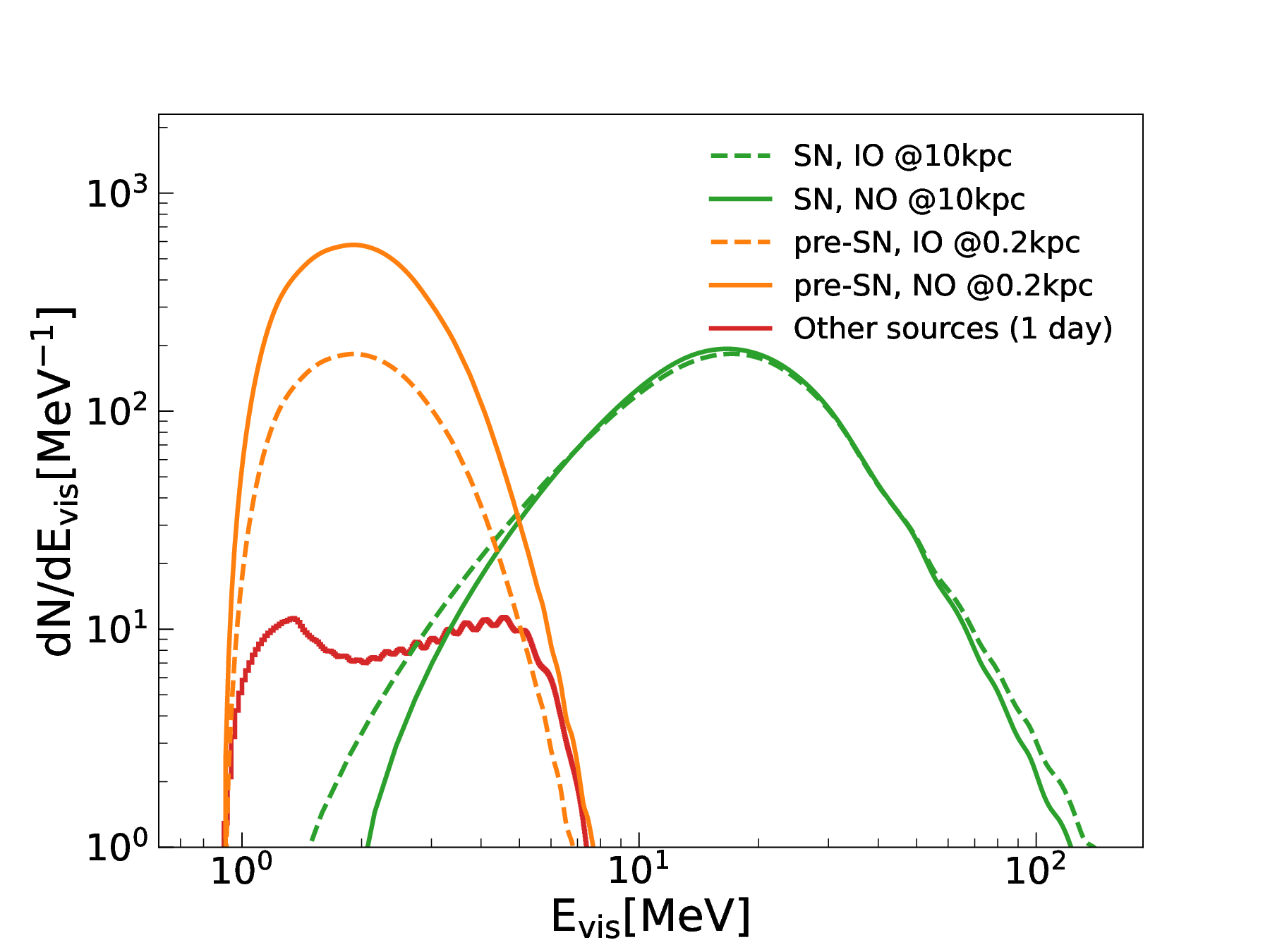}
\end{center}
\vspace{-0.5cm}
\caption{The expected energy spectra of IBD events from SN neutrinos with the Nakazato model at 10 kpc, pre-SN neutrinos integrated in the last day before core collapse with the Patton model at 0.2 kpc and the sum of continuous IBD backgrounds, which mainly come from reactor neutrinos at a distance of 53 km, geo-neutrinos, $\beta$-n decays of cosmogenic $^{9}{\rm Li}/^{8}{\rm He}$, fast neutrons, accidental coincidence and $^{13}{\rm C}(\alpha,n)^{16}{\rm O}$ reaction~\cite{JUNO:2022mxj}.}
\label{fig:online_sigAndBkg}
\end{figure}

In the LS detector of JUNO, the IBD background can be contributed by reactor neutrinos emitted from reactor cores of 26.6 ${\rm GW_{th}}$ thermal power at a distance of about 53 km, geo-neutrinos, $\beta$-n decays of the cosmogenic $^{9}{\rm Li}/^{8}{\rm He}$, fast neutrons, accidental coincidence events and the radiogenic process of $^{13}{\rm C}(\alpha,n)^{16}{\rm O}$ reaction~\cite{JUNO:2022mxj}.  Fig.~\ref{fig:online_sigAndBkg} shows the energy spectra of IBDs from SN neutrinos, pre-SN neutrinos and the backgrounds.
The energy range of backgrounds and IBDs of pre-SN neutrinos are the same. So when selecting $pre$IBD candidates, the upper boundary of the prompt energy cut $E_{p}<3.4$~MeV is optimized by maxizing the signal-to-background ratio using pre-SN neutrinos of the 15 $M_{\odot}$ Patton model in the last day before core collapse. The residual backgrounds of the $pre$IBD candidates are about 21/day in total, as summarized in Tab.~\ref{tab:summary}. It is contributed by reactor neutrinos, $^{9}$Li/$^{8}$He, geo-neutrinos and others, with the yield being 18.2, 0.2, 1.2 and 1.4 per day respectively. And in terms of the $sn$IBD candidates, the total background rate is estimated to be about 39/day (13/day) for $E_{\rm low}=5.5$~MeV ($E_{\rm low}=7.5$~MeV), as summarized in Tab.~\ref{tab:summary}. The breakdown of this background rate for each source is 8.3/day (0.4/day) from reactor neutrinos, 30.6/day (12.9/day) from $^{9}$Li/$^{8}$He and 0.3/day (0.1/day) from other sources. The difference of background contribution between $pre$IBD and $sn$IBD stems from the different energy range and muon veto strategy.

Once the $pre$IBD and $sn$IBD candidates are selected, the \textit{Event Accumulator} will keep all of them for the online CCSN characterization, together with other possible events according to the specific requirements.
Once a critical value of the candidate event rate is reached in the \textit{Pre-SN monitor} or \textit{SN Monitor}, an alert will be sent to the \textit{Alert Processer}. At the \textit{Alert Processor}, there are three defined types of alert status. The first type is a SN alert, which is triggered when the \textit{SN monitor} detects an alert. The second type is a pre-SN alert, which is triggered when the \textit{Pre-SN monitor} detects an alert. The third type is a nearby alert, which is triggered when both the \textit{SN monitor} and \textit{Pre-SN monitor} detect alerts within a certain time frame, such as 10 days.

{The \textit{Fast Characterization} is designed to provide in time characterization of the CCSN candidate, e.g., the CCSN direction, energy spectra and light curves with the input events from the \textit{Event Accumulator}.}
Here we take the CCSN direction reconstruction (cf. Sec.~\ref{sec:dir}) as illustration to present the basic procedures.
In case of an SN alert or nearby alert, a time window of 20 s is opened from 1 s before to 19 s after the alert timestamp. All the $sn$IBD events within this time window are combined to reconstruct the CCSN direction. In particular for a nearby alert, which indicates that a nearby Galactic CCSN may have occurred and a large amount of SN candidate events would flood into the DAQ system, the challenges to the DAQ computing ability should also be carefully treated. Therefore it is possible that the \textit{Event Reconstruction} and its subsequent processes are stopped by DAQ to handle such a specific case, and the online monitor will not be able to reconstruct CCSN direction based on SN neutrino signals itself. Fortunately, even in this extreme case there is still an opportunity to extract the direction of the nearby CCSN if a certain number of $pre$IBD events are available. The reconstructed direction will be updated as more $pre$IBD are accumulated.
All the reconstructed direction information will be sent to the onsite computer and further aid the telescopes to constrain the sky coverage. In the end, the \textit{Alert Processer} will inform the DAQ manager in case of an SN alert or nearby alert in order to store the raw triggerless T/Q data within a predefined time interval (e.g. $\pm$60 s). The storage of the normal data streams is not affected at the same time. Such a design can limit the loss of unforeseen information of the CCSN and preserve extra potential in the future.

\section{Sensitivity of the CCSN Early Alerts}
\label{sec:alert}

In the real-time monitoring system, an intrinsic problem of detecting a CCSN candidate is to identify the transient variation of the event rates induced by the pre-SN neutrinos or SN neutrinos.
The background rates of the candidate signals are relatively stable and could be precisely determined and updated by \textit{in-situ} measurements.
To search for an event rate increase compared to the background,
we apply a \textit{sliding event} method where the time interval $\Delta T$ of the $N$ latest sequential candidates is calculated. 
If the $\Delta T$ is smaller than a predefined time interval threshold $\Delta T_{\rm thr}$, a CCSN candidate is identified and the alert will be released.
This \textit{sliding event} method is equivalent to the typical \textit{sliding window} method, which counts the number of events within a fixed time window, but the \textit{sliding event} method provides a finer granularity to tune the alert system when the number of events N is small.

The choice of $\Delta T_{\rm thr}$ and $N$ in the \textit{sliding event} method depends on the time duration of pre-SN neutrinos or SN neutrinos and the alert performance of the monitoring system, which are generally quantified using four parameters:
\begin{itemize}
\item The \textit{alert efficiency} of correctly releasing a CCSN alert with pre-SN neutrinos or SN neutrinos. It can be estimated using Monte Carlo samples, calculating the fraction of samples that give alerts.
\item  The \textit{alert distance} at where the \textit{alert efficiency} reaches 50\%. 
\item The \textit{alert time} that the monitoring system releases an CCSN alert. In the following, the \textit{alert time} is evaluated to be the time relative to the collapse time for pre-SN neutrinos, and the bounce time for SN neutrinos. In reality, the time latency due to the data transfer, the candidate selection and criteria tests is estimated to be around $O(1)$~s for the prompt monitor and $O(1)$ minute for the online monitor.
\item The \textit{false alert rate} (FAR) induced by the background statistical fluctuation. Note that choosing parameters that increases the SN coverage and reduces the alert time, would simultaneously increase the FAR.
\end{itemize}


In the following part of this section, Monte Carlo simulations of pre-SN neutrinos and SN neutrinos are performed with the JUNO official framework~\cite{JUNO:2021vlw} to evaluate the capability of the monitoring system for the CCSN. The simulation includes a full chain of the event generator, detector simulation, electronics simulation, trigger algorithm and event reconstruction. The time series of background candidates are produced separately in a toy Monte Carlo manner and mixed with the CCSN-related events during the monitoring procedure. The FAR under different sets of $\Delta T_{\rm thr}$ and $N$ are quantified using random background candidates with the proper Poisson fluctuation for the online monitor. However, note that in the prompt monitor case the background mainly comes from the cosmogenic isotopes, whose multiplicity might be larger than one in a single muon. Therefore, both the multiple production of the isotopes in a single muon and the random coincidence of these single events from different muons should be considered when one estimate the false alert rate of the CCSN in the selected time window. Unexpected transient signals (i.e. possible PMT flashers) that may cause false alert will be considered in the future. In the following, the settings will be chosen so the conservative FAR baseline of 1/month and 1/year are adopted for the JUNO monitoring system. As already mentioned before, the prompt monitor will not be used for the pre-SN monitor because of the much lower event rate and neutrino energies. However, since the online monitor employs the reconstructed information and event selection, it is applicable for both pre-SN and SN monitors.

\subsection{Sensitivity of the Prompt Monitor}
Following the logic of the prompt monitor on the global-trigger board as discussed in Section \ref{sec:Monitor}, the sensitivity of the prompt monitor to SN neutrinos is evaluated with the global trigger system. As discussed in Section \ref{sec:PMgt}, different values of $N_{\rm low}$ are used to meet the requirement of having a FAR smaller than 1/month or 1/year. The corresponding background rates are estimated to be 83/day and 45/day, respectively.
\begin{figure}
\begin{center}
\begin{tabular}{l}
\includegraphics[width=0.5\textwidth]{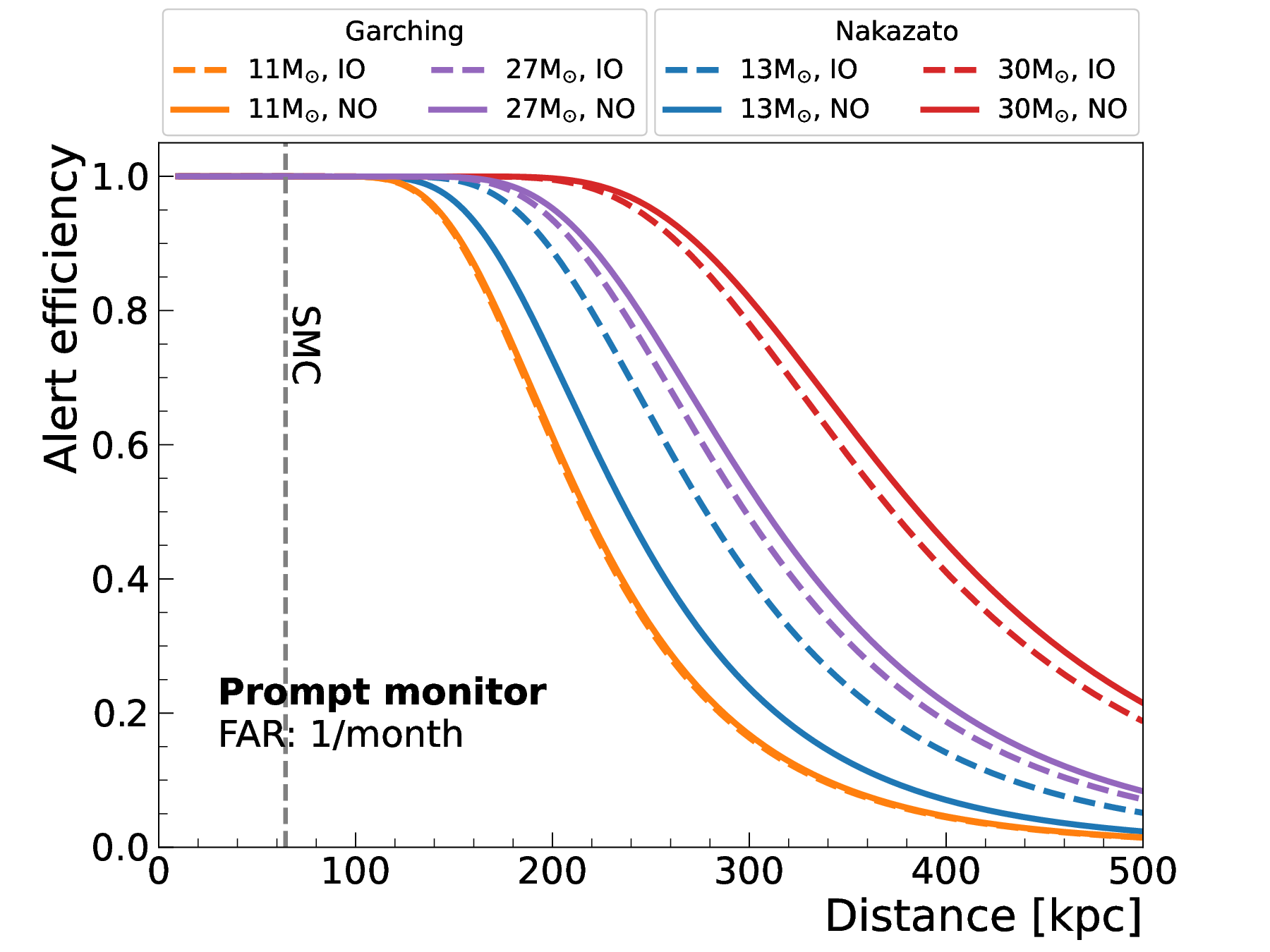}
\hspace{-0.8cm}
\includegraphics[width=0.5\textwidth]{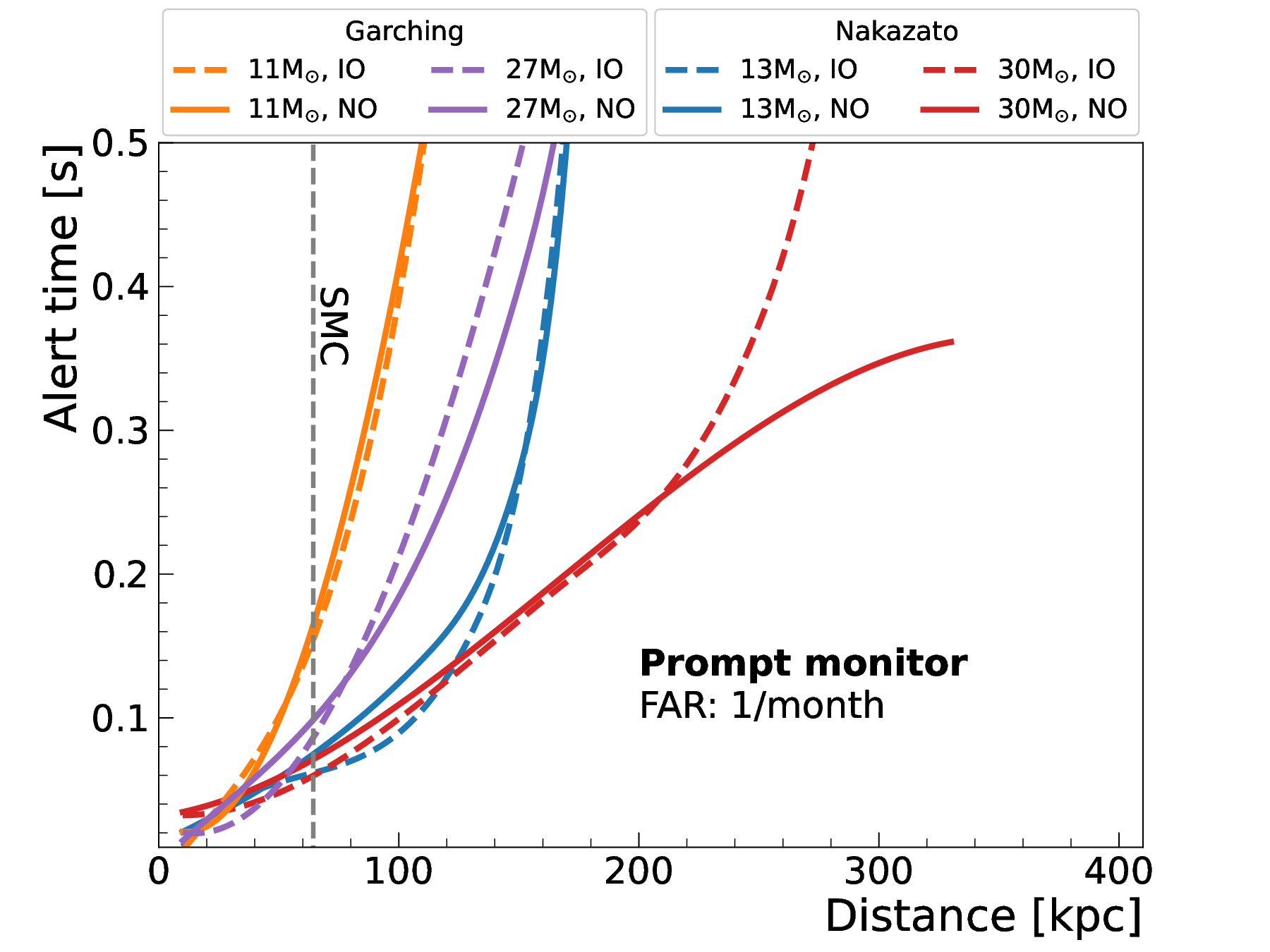} \\
\includegraphics[width=0.5\textwidth]{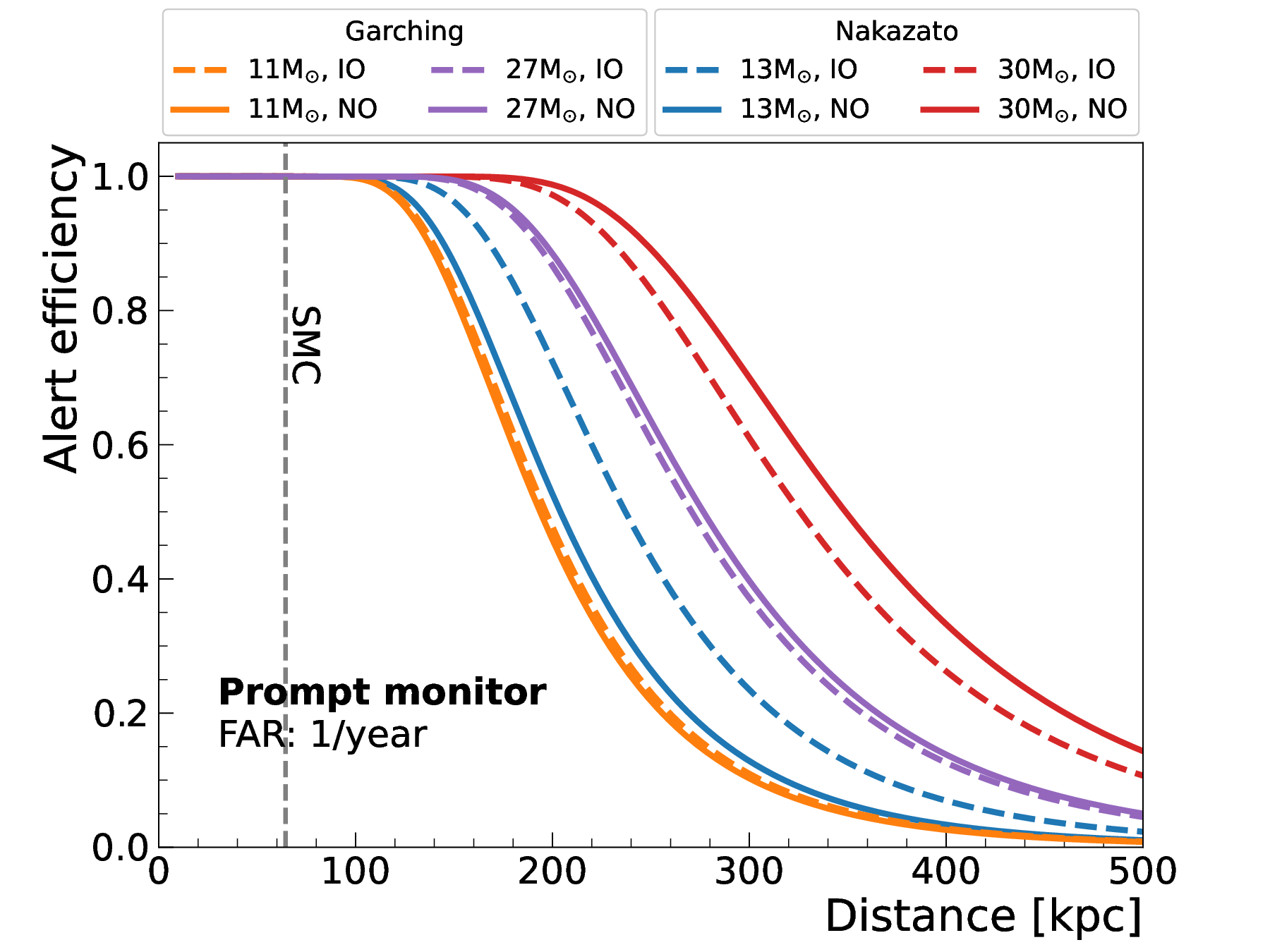}
\hspace{-0.8cm}
\includegraphics[width=0.5\textwidth]{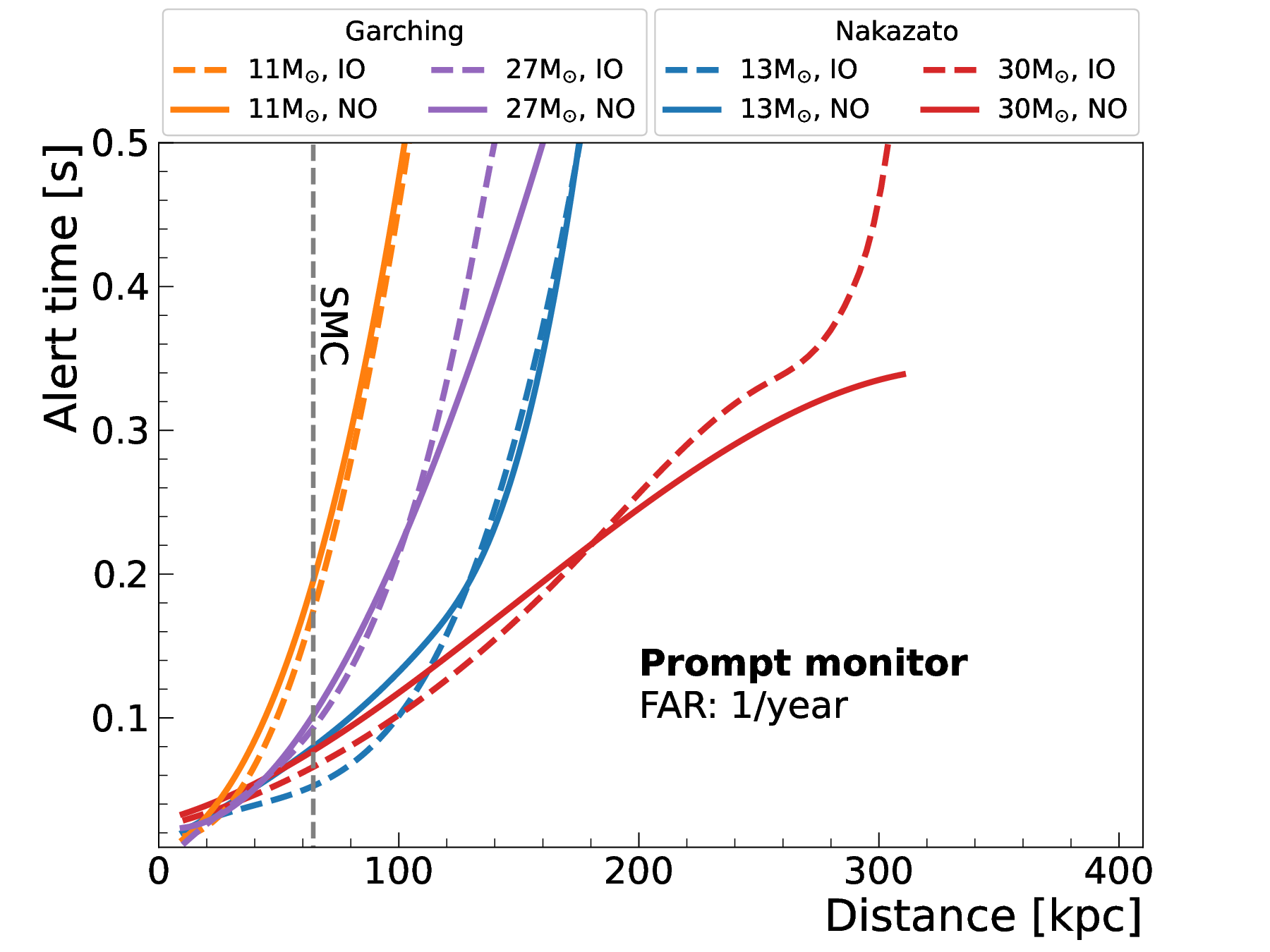}
\vspace{-0.5cm}
\end{tabular}
\end{center}
\caption{Expected alert efficiency (left panel) and alert time relative to the bounce moment (right panel) for the prompt monitor with SN neutrinos of Nakazato and Garching models at different distances. 
FAR of 1/month (top row) and 1/year (bottom row) are assumed.}
\label{fig:alertEffTime}
\end{figure}
The performance of the prompt monitor is evaluated with simulated SN neutrino events from both the Nakazato and Garching models, {whose information is described in Sec.~\ref{sec:neutrinos}.} In the \textit{sliding event} method of the prompt monitor, the time interval threshold $\Delta T_{\rm thr}$ is determined with a fixed number of event candidates $N=3$ and is set to 7.5 s and 2.1 s for a FAR of 1/month and 1/year, respectively. 
The results on the alert efficiency and alert time for different models are shown in Fig.~\ref{fig:alertEffTime}, while the alert distance 
{and alert time at $10$ kpc} of the prompt monitor is summarized in the right part of Tab.~\ref{tab:summary}. The alert time after the core bounce at each distance is defined as the 68\% upper bound of the issued alert time for all tested samples.
As shown in Tab.~\ref{tab:summary} and the right panel of Fig.~\ref{fig:alertEffTime}, the intrinsic alert time after bounce is about $20\sim30$~ms for the Nakazato models and $10\sim20$ ms for the Garching models for a CCSN with the distance of 10 kpc. At this distance, the alert efficiency is 100\% and the alert time is determined by the event rate at early stage. So usually, CCSN progenitors with larger mass and IO will have higher event rate at early stage (e.g. the Nakazato models shown in the right plot of Fig.~\ref{fig:IBDt}) and hence be alerted earlier. {Also note that despite a larger overall signal in the IO case for 13~$M_\odot$, the time profile of the emitted events will impact the efficiency to identify those events, which is why the NO scenario shows a greater efficiency than the IO scenario.}
\begin{sidewaystable}
\centering
\footnotesize
\begin{tabular}{c|c|c|c|c|c|c|cc|cc}
\hline
\multirow{2}{*}{} & \multirow{2}{*}{Model} & \multirow{2}{*}{\begin{tabular}[c]{@{}c@{}}Mass\\ {[}$M_{\odot}${]}\end{tabular}} & \multirow{2}{*}{\begin{tabular}[c]{@{}c@{}}Mass\\ ordering\end{tabular}} & \multirow{2}{*}{\begin{tabular}[c]{@{}c@{}}$r_{bkg}$\\ {[}$day^{-1}${]}\end{tabular}} & \multirow{2}{*}{$N_{IBD}$} & \multirow{2}{*}{$N_{sel}$} & \multicolumn{2}{c|}{Alert distance {[}kpc{]}} & \multicolumn{2}{c}{Alert time} \\ \cline{8-11} 
 &  &  &  &  &  &  & \multicolumn{1}{c|}{FAR\textless{}1/month} & FAR\textless{}1/year & \multicolumn{1}{c|}{FAR\textless{}1/month} & FAR\textless{}1/year \\ \hline
\multirow{16}{*}{SN} & \multirow{8}{*}{Garching} & \multirow{4}{*}{11} & \multirow{2}{*}{NO} & \multirow{16}{*}{\begin{tabular}[c]{@{}c@{}}39\\ (83)\end{tabular}} & \multirow{2}{*}{1675} & 1414 & \multicolumn{1}{c|}{230} & 230 & \multicolumn{1}{c|}{\multirow{2}{*}{(16 ms)}} & \multirow{2}{*}{(17 ms)} \\
 &  &  &  &  &  & (1204) & \multicolumn{1}{c|}{(220)} & (190) & \multicolumn{1}{c|}{} &  \\ \cline{4-4} \cline{6-11} 
 &  &  & \multirow{2}{*}{IO} &  & \multirow{2}{*}{1676} & 1413 & \multicolumn{1}{c|}{230} & 230 & \multicolumn{1}{c|}{\multirow{2}{*}{(13 ms)}} & \multirow{2}{*}{(14 ms)} \\
 &  &  &  &  &  & (1228) & \multicolumn{1}{c|}{(220)} & (200) & \multicolumn{1}{c|}{} &  \\ \cline{3-4} \cline{6-11} 
 &  & \multirow{4}{*}{27} & \multirow{2}{*}{NO} &  & \multirow{2}{*}{3132} & 2651 & \multicolumn{1}{c|}{320} & 320 & \multicolumn{1}{c|}{\multirow{2}{*}{(15 ms)}} & \multirow{2}{*}{(16 ms)} \\
 &  &  &  &  &  & (2466) & \multicolumn{1}{c|}{(310)} & (280) & \multicolumn{1}{c|}{} &  \\ \cline{4-4} \cline{6-11} 
 &  &  & \multirow{2}{*}{IO} &  & \multirow{2}{*}{2958} & 2502 & \multicolumn{1}{c|}{310} & 310 & \multicolumn{1}{c|}{\multirow{2}{*}{(13 ms)}} & \multirow{2}{*}{(13 ms)} \\
 &  &  &  &  &  & (2366) & \multicolumn{1}{c|}{(300)} & (270) & \multicolumn{1}{c|}{} &  \\ \cline{2-4} \cline{6-11} 
 & \multirow{8}{*}{Nakazato} & \multirow{4}{*}{13} & \multirow{2}{*}{NO} &  & \multirow{2}{*}{2326} & 1934 & \multicolumn{1}{c|}{270} & 240 & \multicolumn{1}{c|}{\multirow{2}{*}{(20 ms)}} & \multirow{2}{*}{(21 ms)} \\
 &  &  &  &  &  & (1698) & \multicolumn{1}{c|}{(240)} & (200) & \multicolumn{1}{c|}{} &  \\ \cline{4-4} \cline{6-11} 
 &  &  & \multirow{2}{*}{IO} &  & \multirow{2}{*}{2827} & 2365 & \multicolumn{1}{c|}{300} & 270 & \multicolumn{1}{c|}{\multirow{2}{*}{(16 ms)}} & \multirow{2}{*}{(17 ms)} \\
 &  &  &  &  &  & (2190) & \multicolumn{1}{c|}{(280)} & (240) & \multicolumn{1}{c|}{} &  \\ \cline{3-4} \cline{6-11} 
 &  & \multirow{4}{*}{30} & \multirow{2}{*}{NO} &  & \multirow{2}{*}{5074} & 4098 & \multicolumn{1}{c|}{400} & 370 & \multicolumn{1}{c|}{\multirow{2}{*}{(31 ms)}} & \multirow{2}{*}{(31 ms)} \\
 &  &  &  &  &  & (4217) & \multicolumn{1}{c|}{(390)} & (350) & \multicolumn{1}{c|}{} &  \\ \cline{4-4} \cline{6-11} 
 &  &  & \multirow{2}{*}{IO} &  & \multirow{2}{*}{4972} & 4131 & \multicolumn{1}{c|}{390} & 350 & \multicolumn{1}{c|}{\multirow{2}{*}{(31 ms)}} & \multirow{2}{*}{(31 ms)} \\
 &  &  &  &  &  & (4145) & \multicolumn{1}{c|}{(370)} & (330) & \multicolumn{1}{c|}{} &  \\ \hline
\multirow{4}{*}{pre-SN} & \multirow{4}{*}{Patton} & \multirow{2}{*}{15} & NO & \multirow{4}{*}{21} & 659 & 556 & \multicolumn{1}{c|}{1.3} & 1.1 & \multicolumn{1}{c|}{-140 h} & -120 h \\ \cline{4-4} \cline{6-11} 
 &  &  & IO &  & 196 & 156 & \multicolumn{1}{c|}{0.7} & 0.6 & \multicolumn{1}{c|}{-90 h} & -30 h \\ \cline{3-4} \cline{6-11} 
 &  & \multirow{2}{*}{30} & NO &  & 1176 & 930 & \multicolumn{1}{c|}{1.7} & 1.6 & \multicolumn{1}{c|}{-220 h} & -180 h \\ \cline{4-4} \cline{6-11} 
 &  &  & IO &  & 379 & 302 & \multicolumn{1}{c|}{1.0} & 0.9 & \multicolumn{1}{c|}{-100 h} & -3 h \\ \hline
\end{tabular}
\caption{This table summarizes several quantities related to the online monitor (prompt monitor), including the background rate ($r_{bkg}$), expected number of signals ($N_{IBD}$) and selected candidates ($N_{sel}$), alert distance and alert time for different models. When selecting the signals, $N_{\rm hit}$ is used for the prompt monitor and IBD events with different energy ranges are used for online SN and pre-SN monitor, refer to text for details. And the $N_{IBD}$ is estimated with different models @10kpc for the SN neutrinos and @0.2kpc within the last day before core collapse for the pre-SN neutrinos. $r_{bkg}$ and $N_{sel}$ for prompt monitor and online SN monitor is estimated using the selection criteria corresponding to FAR being 1/month.}
\label{tab:summary}
\end{sidewaystable}

\subsection{Sensitivity of the Online Monitor}

The online monitor has potential sensitivities to both pre-SN neutrinos and SN neutrinos through finely reconstructed and filtered candidates, which could provide early alerts for a nearby Galactic progenitor star before core collapse as well as a Galactic or nearby extragalatic CCSN during explosion. 

\subsubsection{Pre-SN neutrinos}

\begin{figure}
\begin{center}
\begin{tabular}{l}
\includegraphics[scale=0.26]{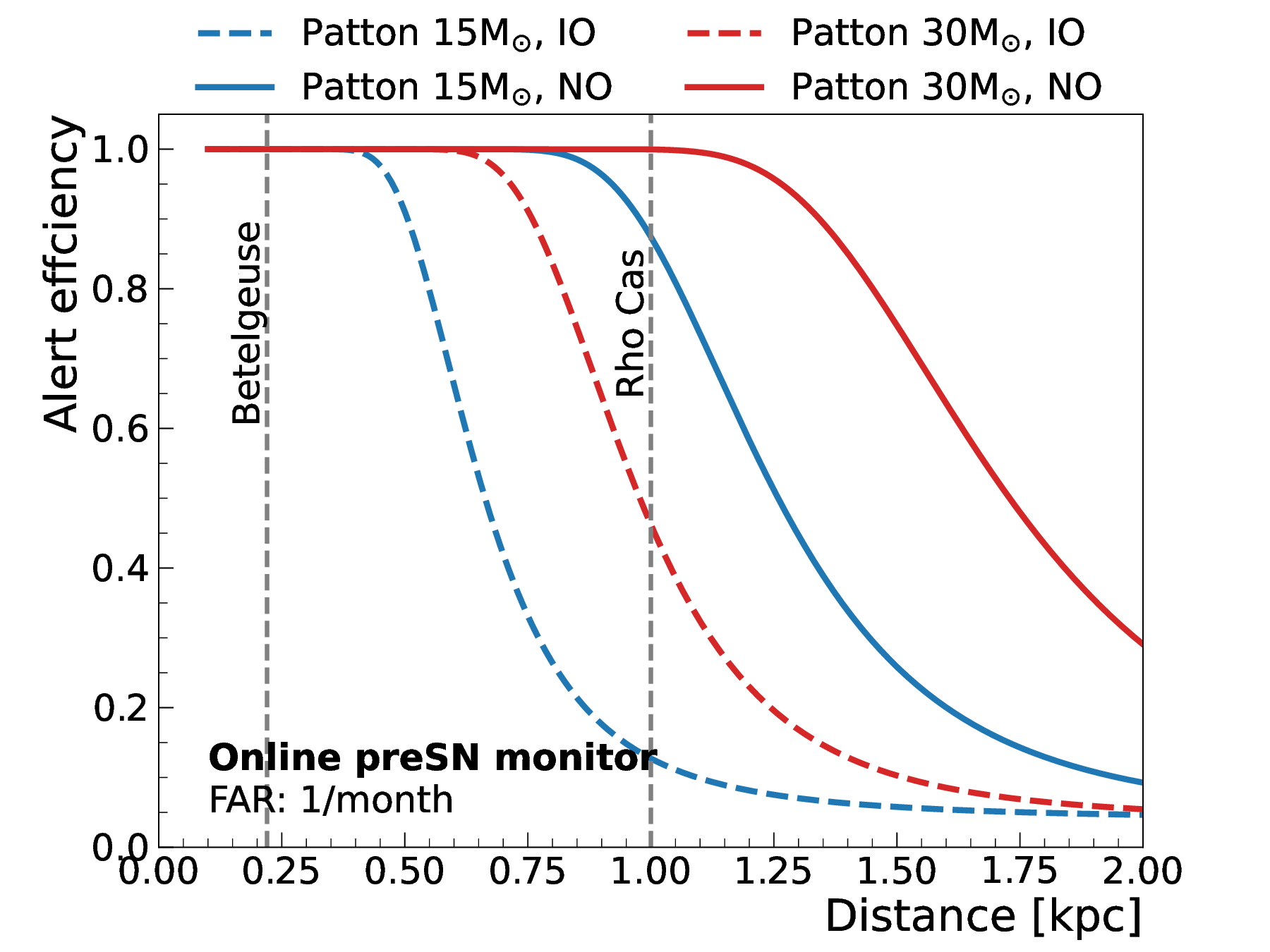}
\includegraphics[scale=0.26]{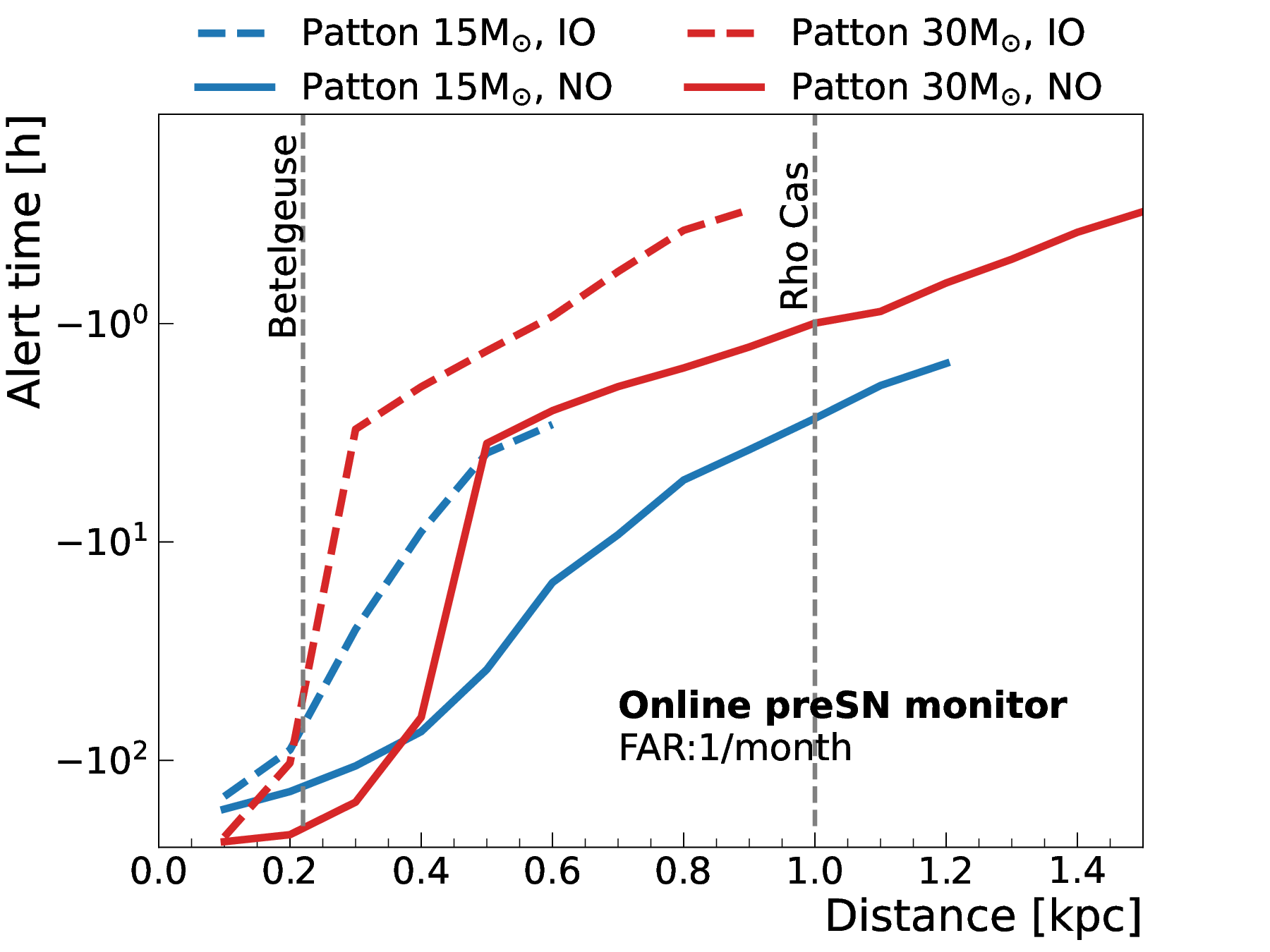} 
\\
\includegraphics[scale=0.26]{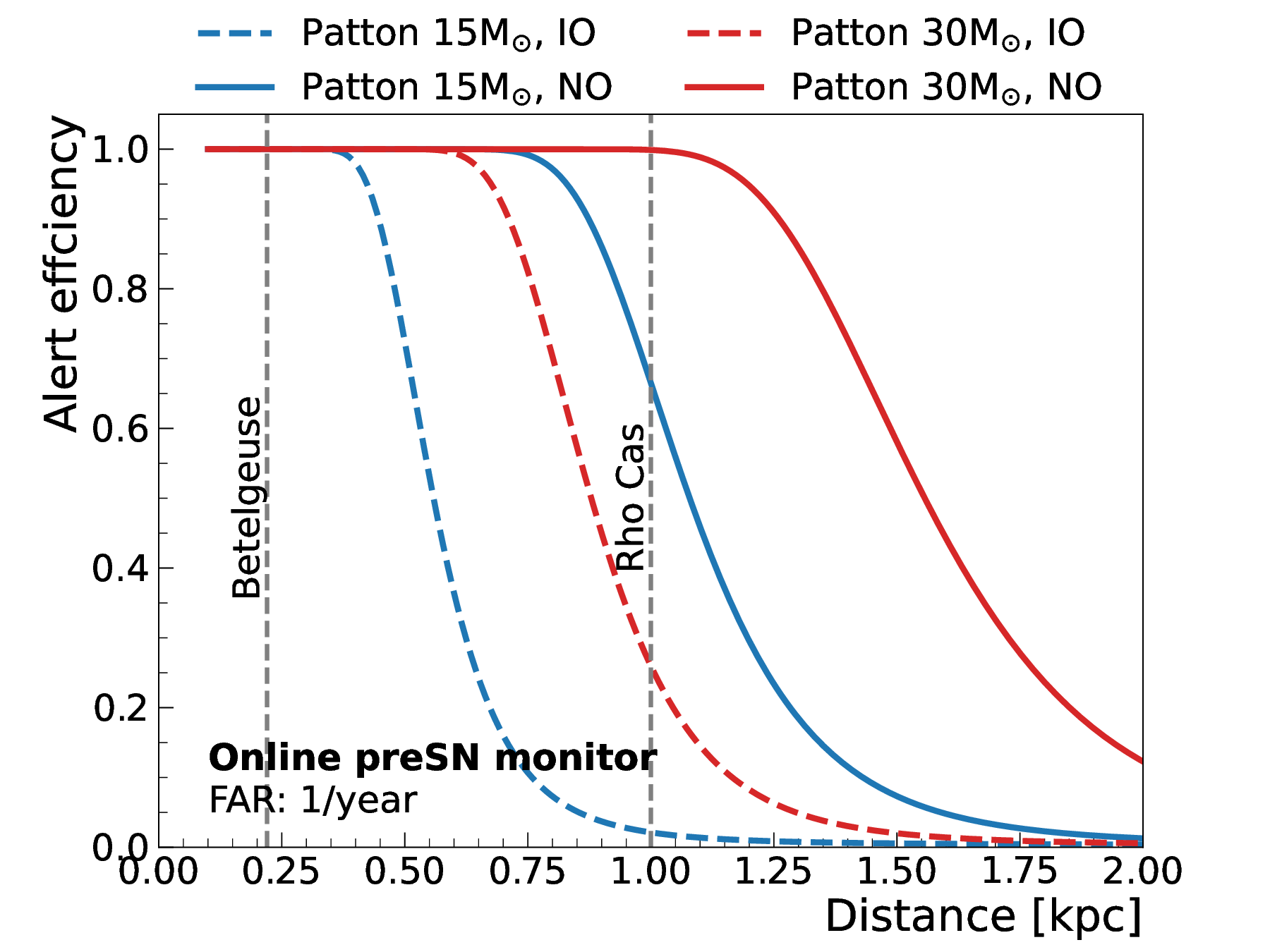}
\includegraphics[scale=0.26]{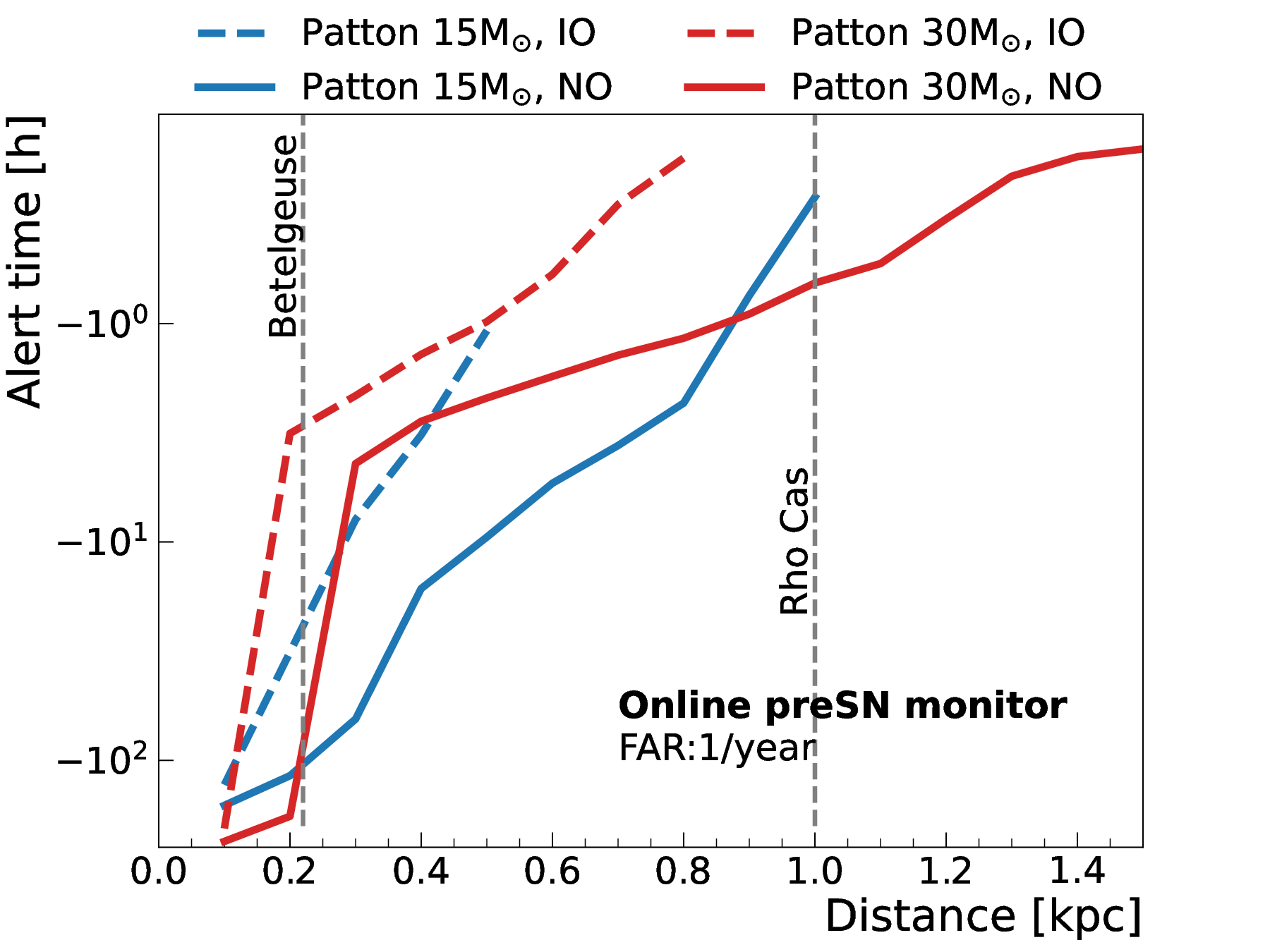}
\end{tabular}
\end{center}
\vspace{-0.5cm}
\caption{Alert efficiency (left panel) and alert time (right panel) of the online \textit{Pre-SN Monitor} at different distances for progenitor stars of $\rm15M_{\odot}$ and $\rm30M_{\odot}$ Patton models for the NO and IO cases. The FAR is 1/month (upper panel) and 1/year (lower panel).}
\label{fig:preSN_eff_time}
\end{figure}

As mentioned in Section \ref{sec:DAQ}, the online monitor can utilize either the pre-SN neutrinos or SN neutrinos for the early warning purpose.
The detection of pre-SN neutrinos can provide an early alert for an imminent CCSN, which can be realized with the monitoring of the $pre$IBD candidates as described in Sec.~\ref{sec:DAQ}.
Sec.~\ref{sec:DAQ} has already discussed the selection of $pre$IBD candidates and its background, whose event rate is estimated to be about 21/day. The performance of the online pre-SN monitor is evaluated with the \textit{sliding event} method, whose parameters are determined in the case of a FAR of 1/month or 1/year. The performance of alert efficiency to nearby progenitor stars is shown in Fig.~\ref{fig:preSN_eff_time} using the Patton models. The corresponding alert distances and alert time at $0.2$~kpc are summarized in the right part of Tab.~\ref{tab:summary}. 
In the worst scenario of IO and the 15 $M_{\odot}$ Patton model, the online pre-SN monitor with a FAR of 1/year would be sensitive to the progenitor stars within 0.6~kpc away from the Earth, where about 22 nearby massive stars are covered according to the list of candidate pre-SN stars of Ref.~\cite{Mukhopadhyay:2020ubs}.

Betelgeuse is often taken as a representative of the nearby massive progenitor stars and a candidate source of pre-SN neutrinos. To compare with pre-SN studies in KamLAND~\cite{KamLAND:2015dbn} and Super-Kamiokande with Gadolinium~\cite{Super-Kamiokande:2019xnm}, we assume two extreme conditions of 0.15 kpc and 15 $M_{\odot}$, and 0.25 kpc and 30 $M_{\odot}$ for Betelgeuse, instead of the up-to-date estimation of the distance and mass as in Ref.~\cite{Betelgeuse:2020}. The expected alert time of a Betelgeuse-like star in the online pre-SN monitor 
{turns out to be 141 hr (84 hr) and 73 hr (2.6 hr) before the core collapse for the $15M_{\odot}$ and $30M_{\odot}$ Patton models with NO (IO) assuming the false alert rate is 1/year.} Even in the worst case of IO and a FAR of 1/year, a Betelgeuse-like star at 0.25~kpc with 30~$M_{\odot}$ will still cause an alert about 2.6 hr before the core collapse.
By utilizing the same pre-SN model and mass ordering option, the alert time of 141 hours prior to the core collapse for JUNO is better than the sensitivity of KamLAND, with a alert time of 89.6 hours (low reactor scenario) or 46.0 hours (high reactor scenario). In addition to that, JUNO exhibits an alert distance of 1.6 kpc, while Super-Kamiokande with Gadolinium achieves an alert distance of 0.6 kpc.
JUNO is expected to surpasses KamLAND in performance due to its 20 times larger target mass, while it is expected to outperform Super-Kamiokande with Gadolinium thanks to the expected improvement in signal-to-background ratio.


\subsubsection{SN neutrinos}
\begin{figure}
\begin{center}
\begin{tabular}{l}
\includegraphics[scale=0.26]{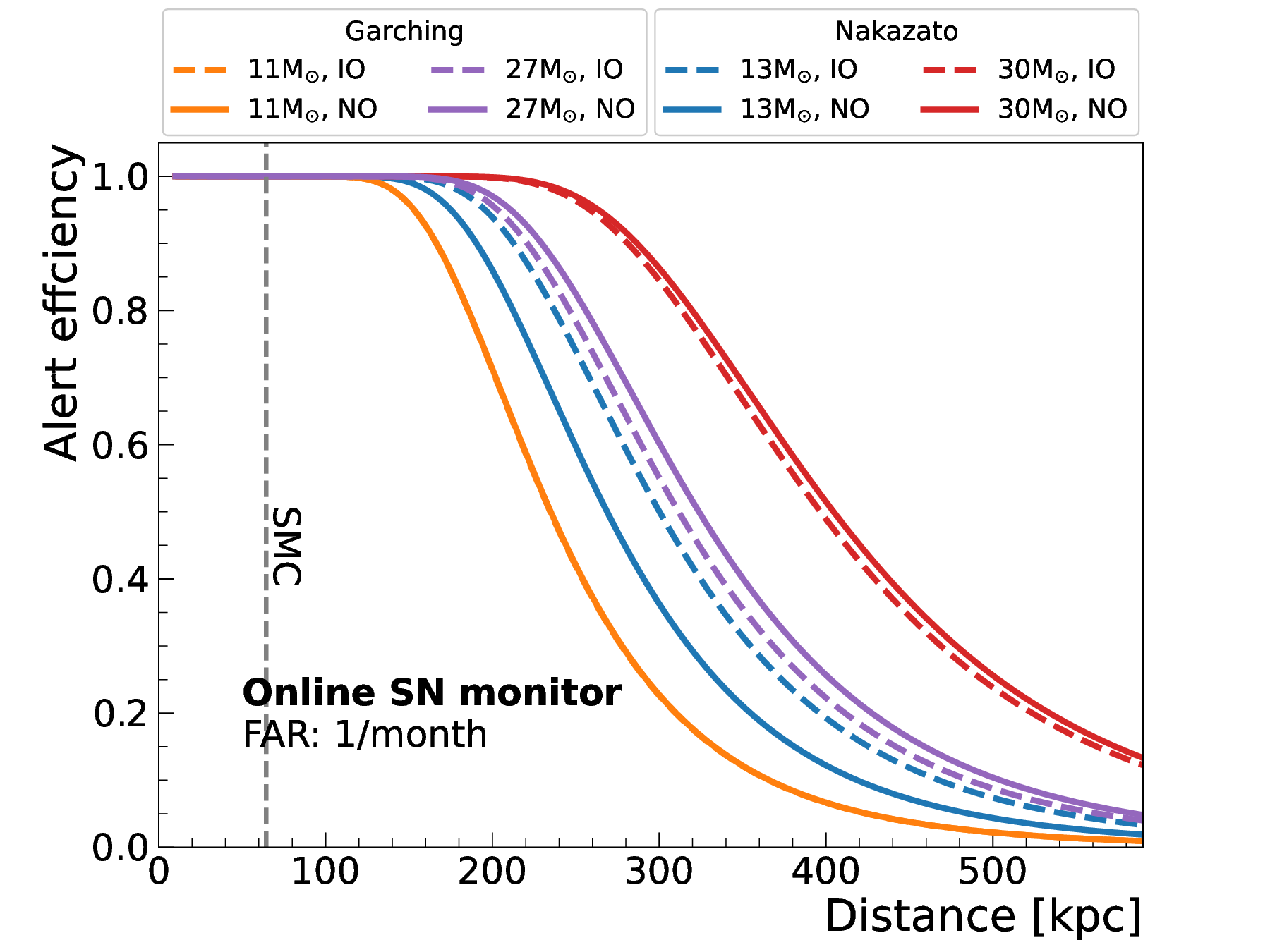}
\includegraphics[scale=0.26]{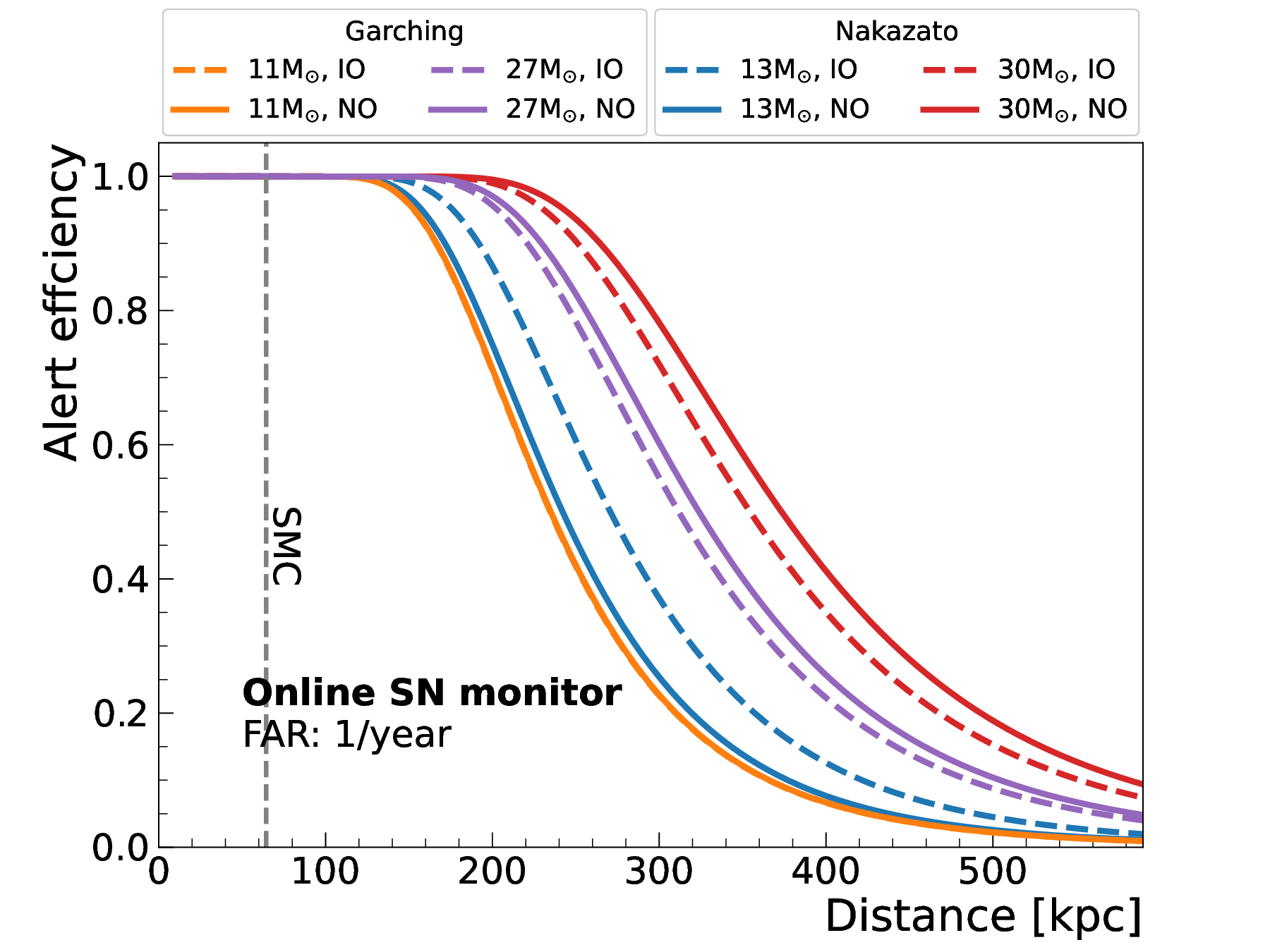}
\end{tabular}
\end{center}
\vspace{-0.5cm}
\caption{Alert efficiency of the online \textit{SN Monitor} for the CCSN at different distances using 13$M_{\odot}$ and 30$M_{\odot}$ Nakazato models for both the NO and IO cases. The FAR in the left plot is 1 per month while is 1 per year in the right plot.}
\label{fig:SN_eff}
\end{figure}

As discussed in Sec.~\ref{sec:DAQ}, different values of $E_{low}$ are used to meet the requirement of having a FAR smaller than 1/month and 1/year. The corresponding background rates are estimated to be 39/day and 13/day, respectively. The performance is also evaluated with the \textit{sliding event} method, whose parameters are determined in the case of a FAR of 1/month or 1/year. However, for the online SN monitor, we shall focus on the performance in terms of the alert efficiency rather than the alert time. Because due to the large amount of events accumulated over a short time, the time latency of a SN alert in the online monitor is dominated by the DAQ processing speed, which is anticipated to be at the minute level and will be evaluated with future \textit{in situ} measurements.
The alert efficiency is shown in Fig.~\ref{fig:SN_eff} for the Nakazato and Garching models. A set of parameters $N$ and $T_{\rm thr}$ are chosen as 3 and 15.8~s for $\rm FAR<1/month$ (left panel) and 3 and 4.5~s for $\rm FAR<1/year$ (right panel) in the \textit{sliding event} method. The alert distance is also summarized in the right part of Tab.~\ref{tab:summary}. 
For the Garching models, the alert efficiency are the same when constraining the FAR from 1/month to 1/year, 
which is due to the limited time duration (about 3 to 4 s) of the available data of these models. 
From the figure, one can observe that the online monitor can be sensitive to the CCSN within 230~kpc even in the worst case of the current adopted models, which covers the Milky Way and 51 nearby galaxies according to the catalog of satellite galaxies in Ref.~\cite{Karachentsev:2013cva}.

\section{CCSN pointing}\label{sec:dir}

The CCSN directionality is crucial to guide the telescopes to catch the early light of the CCSN by focusing on the targeted sky area. JUNO will not only contribute to the early alert of the CCSN, which can be used for triangulation pointing along with other worldwide neutrino experiments~\cite{Linzer:2019swe}, but also provide an individual CCSN direction by using the collective anisotropy of the accumulated IBD events from pre-SN neutrinos or SN neutrinos at the DAQ stage.

As studied in Refs.~\cite{Vogel:1999zy,CHOOZ:1999hgz}, IBD events in the LS maintain a moderate direction information of the incoming neutrinos. The neutrons of the IBD events are emitted in a slightly forward angle relative to the neutrino direction and undergo nearly
isotropic scattering and diffusion processes before being captured on hydrogen. The positrons of the IBD events are assumed to annihilate with electrons at their production vertex.
This is a reasonable assumption considering that the average displacement of $O(0.1)$ cm is much smaller than the positron reconstructed vertex resolution~\cite{Qian:2021vnh}. Therefore the direction of the CCSN could be reconstructed using the collective anisotropy of the IBD events from SN or pre-SN neutrinos:
\begin{eqnarray}
\vec{d} = \frac{1}{N} \sum_{i = 1}^{N} {\vec{X}^{i}_{\rm np}} \; ,
\label{eq:dir}
\end{eqnarray}
where $N$ is the total number of IBD events from pre-SN neutrinos or SN neutrinos, $\vec{X}^i_{\rm np}$ (for $i = 1, 2, \cdots, N$) denote the unit vector from the reconstructed neutron vertex to the reconstructed positron vertex for each IBD event. Obviously, the accuracy of the reconstructed CCSN direction is statistics-dependent. In the following evaluation of the pointing ability using full-chain simulated events of the pre-SN neutrinos or SN neutrinos, we use the half-aperture $\theta_{0.68}$ of the cone around the true CCSN direction, which contains 68$\%$ of all sorted results, as the uncertainty of CCSN pointing. The direction reconstruction will be part of the online fast characterization of the CCSN following the logic discussed in Sec.~\ref{sec:alert}.

\begin{figure}
    \centering
    \includegraphics[scale=0.24]{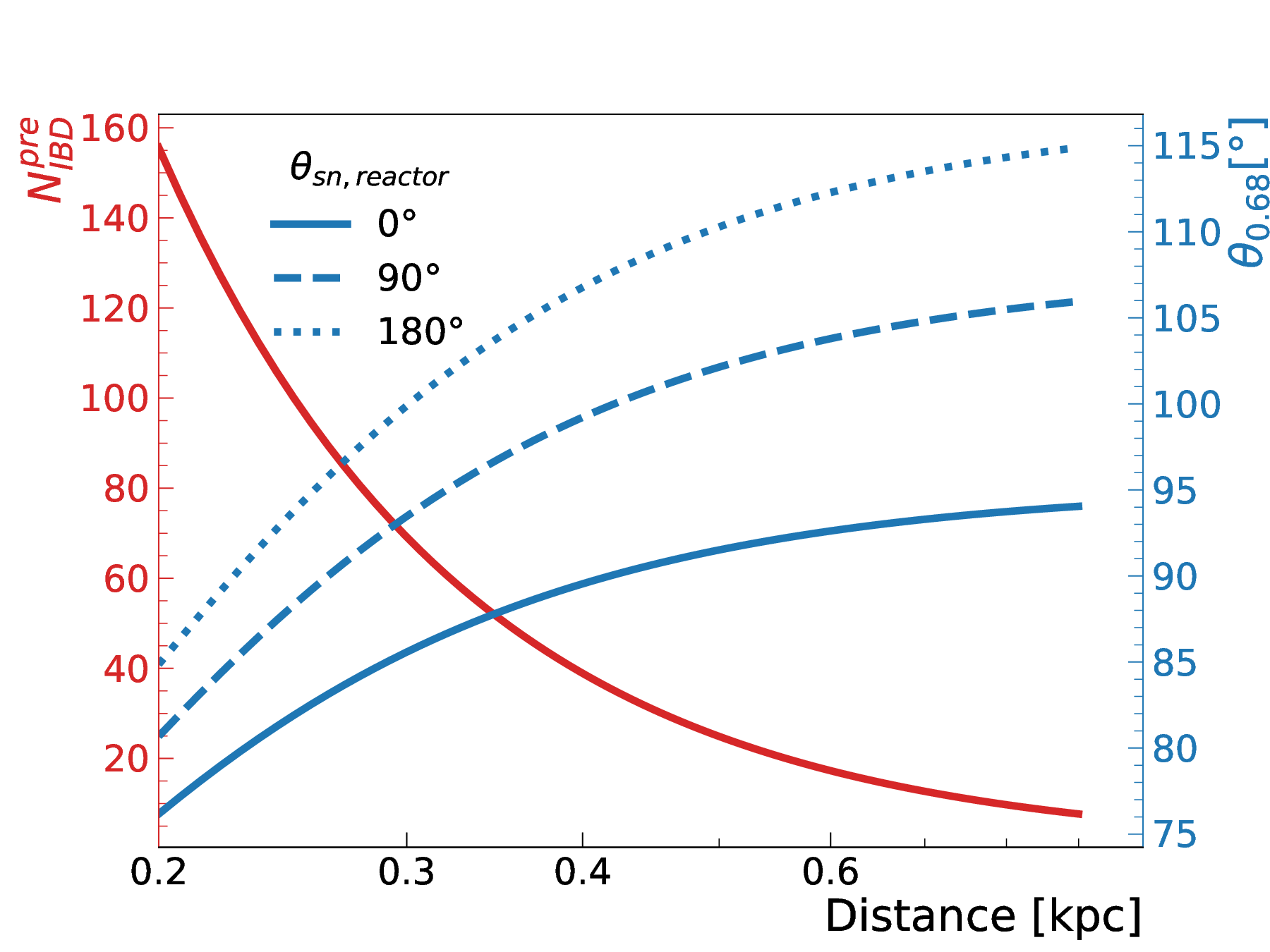}
    \includegraphics[scale=0.24]{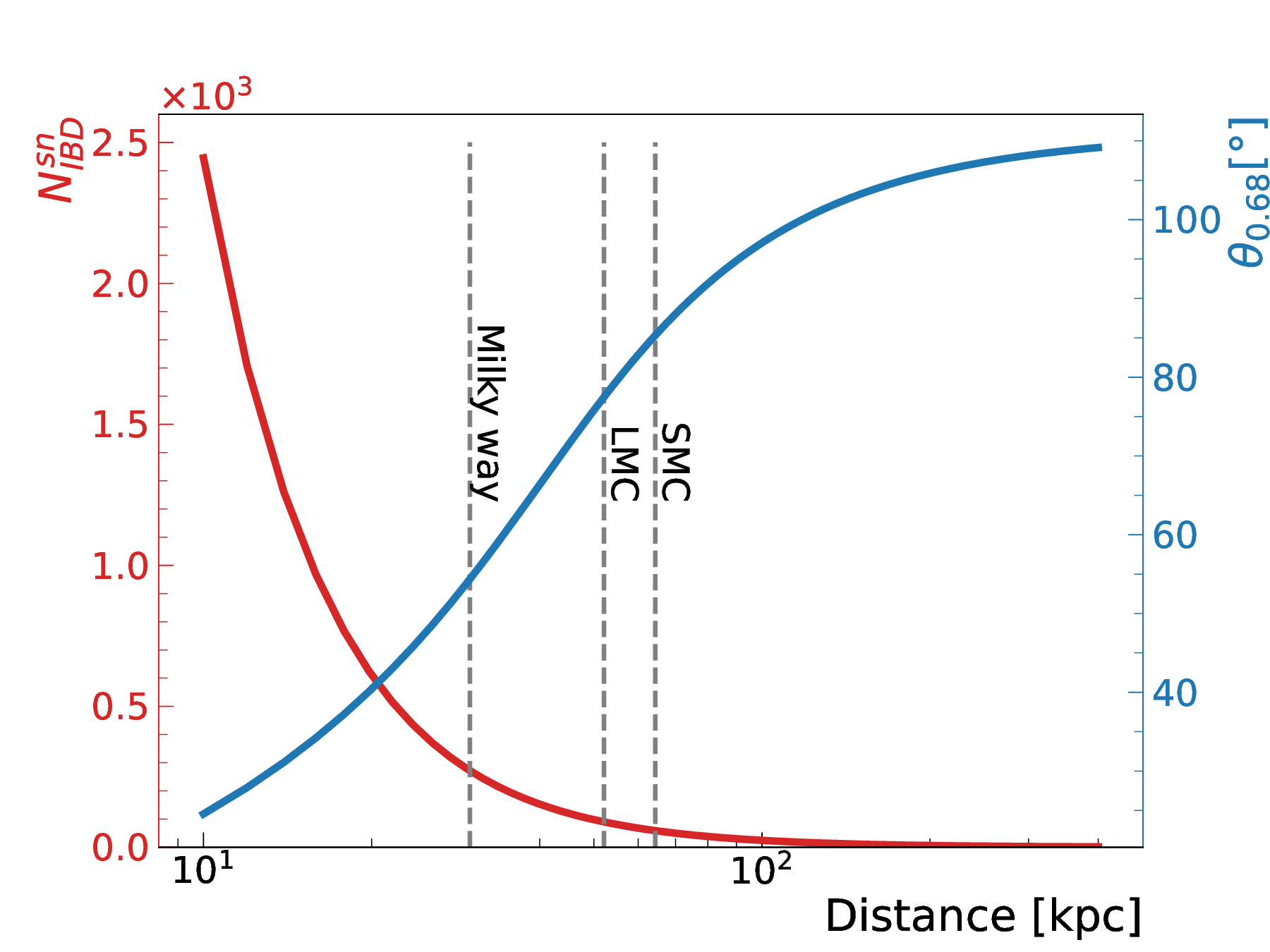}
    \caption{The number of $pre$IBD (left panel) or $sn$IBD (right panel) candidates and the respective CCSN pointing ability as functions of the progenitor distance. In both figures, the number of IBDs is on the left axis and the half-aperture is on the right axis. Note that in the left panel, 0.2~kpc is the distance to Betelgeuse. Also, the pointing ability depends on the number of IBDs, which is scaled to the distance of progenitor stars using 15~$M_{\odot}$, IO Patton model for $pre$IBDs (number of IBD candidates is integrated over the last day before core collapse) and 13~$M_{\odot}$, IO Nakazato model for $sn$IBDs.}
    \label{fig:direction_pre}
\end{figure}

\subsection{Pre-SN Pointing}
Due to the relatively small statistics of $pre$IBD events within several days for a nearby progenitor star, the contribution of the backgrounds in the direction reconstruction should not be neglected, among which the IBD events from reactor neutrinos play the dominant role. Within one day, the relative direction angle $\theta_{\rm sn, reactor}$ between pre-SN neutrinos and reactor neutrinos varies with time due to the rotation of the Earth. Therefore three cases at $\theta_{\rm sn, reactor}=0^{\circ}$, $\theta_{\rm sn, reactor}=90^{\circ}$ and $\theta_{\rm sn, reactor}=180^{\circ}$ are taken into account, where the performance of $\theta_{\rm sn, reactor}=0^{\circ}$ is expected to be the best one and the case of $\theta_{\rm sn, reactor}=90^{\circ}$ somehow corresponds to a realistic average of the pre-SN neutrinos changing direction during the day due to the rotation of Earth. Given a fixed number of background candidates as 21 per day, the reconstructed direction uncertainty varies with the number of $pre$IBD events in the three cases as shown in the left plot of Fig.~\ref{fig:direction_pre}. As the distance increases, there will be less $pre$IBD events and the performance of direction reconstruction becomes worse. In practice, the pointing ability of the pre-SN should be between those in the cases of $\theta_{\rm sn, reactor}=180^{\circ}$ and $\theta_{\rm sn, reactor}=0^{\circ}$ and we use the $\theta_{\rm sn, reactor}=90^{\circ}$ case as a reference. A more detailed reconstruction that accounts for the real direction of pre-SN neutrinos during the observation time would still need to be implemented in the future.

\begin{figure}
\begin{center}
\vspace{-1cm}
\begin{tabular}{l}
\includegraphics[scale=0.4]{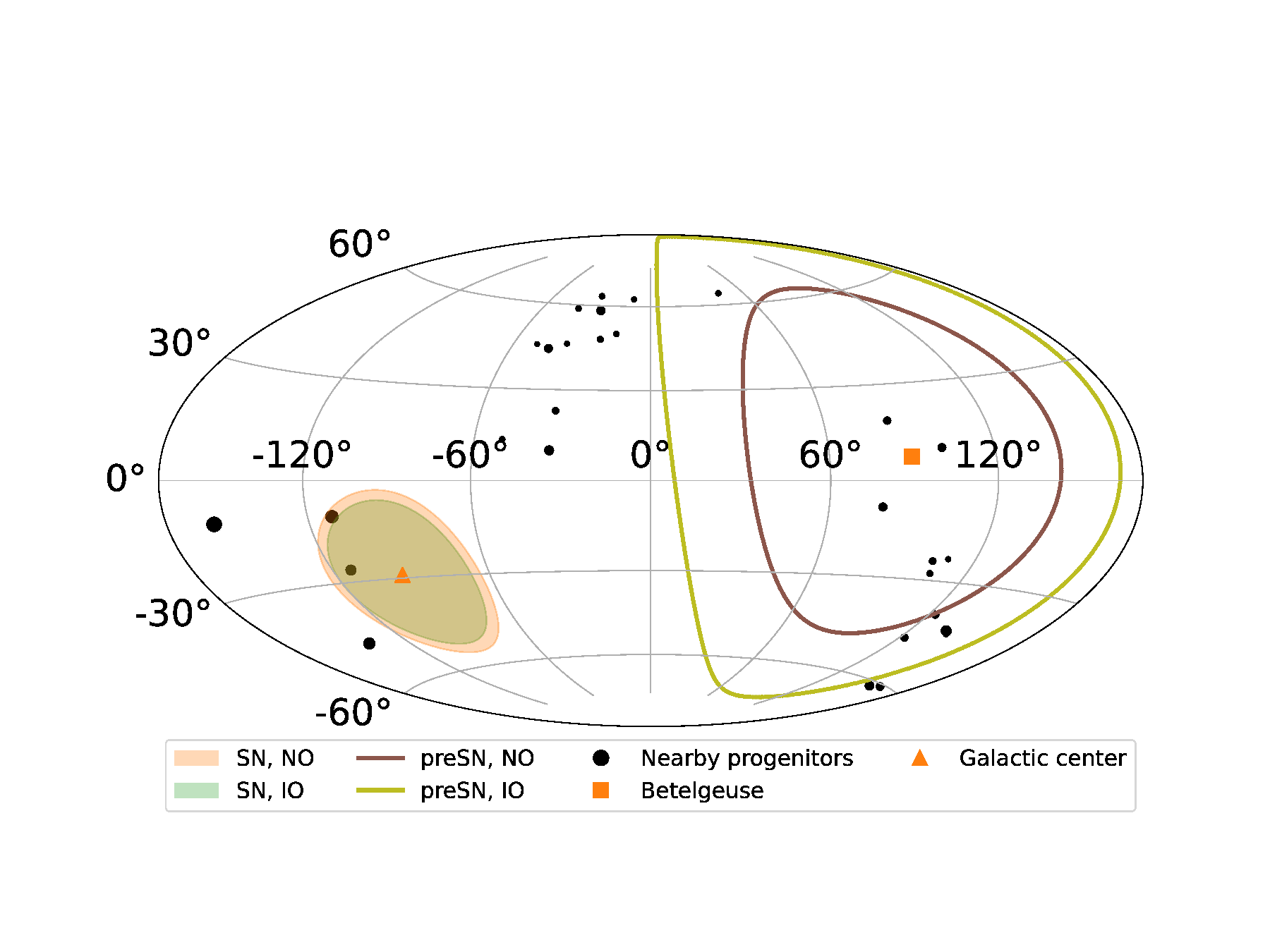}
\end{tabular}
\end{center}
\vspace{-1cm}
\caption{This figure is a sky-map that shows the pointing resolution for pre-SN and SN cases: The red and blue supergiant progenitors as candidates of CCSNe within 1~kpc in the equatorial coordinate system are shown by black dots. The brown and yellow curves circle the 68\% regions of the reconstructed direction for Betelgeuse pre-SN star at 0.2 kpc with NO and IO respectively. The shadowed areas surrounded with the orange and green lines show the pointing precision for a SN at 10~kpc with NO and IO respectively.}
\label{fig:pre_candidate}
\end{figure}

According to the collection of pre-SN progenitor candidates in Ref.~\cite{Mukhopadhyay:2020ubs}, there are 31 red and blue CCSN progenitors within 1 kpc, which is shown as the blue points of Fig.~\ref{fig:pre_candidate} in the equatorial coordinate system (J2000). Taking the well-known progenitor star Betelgeuse as an example, we assume that pre-SN neutrinos are produced according to the $15 M_{\odot}$ Patton model at 0.2 kpc. Following the logic of the online pre-SN monitor, there are 556 (156) $pre$IBD events in the last one day before core collapse in the NO (IO) case. Therefore, the pre-SN pointing with an uncertainty of $56^{\circ}$ ($81^{\circ}$), estimated from the moderate case of $\theta_{\rm sn, reactor}=90^{\circ}$ in Fig.~\ref{fig:direction_pre}, is expected for the NO (IO) case, as illustrated by the area surrounded with the red (yellow) line in Fig.~\ref{fig:pre_candidate}. Such direction information can aid the astronomers to constrain CCSN progenitors at the very last stages of the stellar evolution.

\subsection{SN Pointing}

The background contribution in the CCSN direction reconstruction with the $sn$IBD events is negligible over a short time scale. In the right plot of Fig.~\ref{fig:direction_pre} we illustrate the number of the $sn$IBD events and the corresponding pointing ability with the 13$M_{\odot}$ Nakazato model and IO at different distances.
Assuming a typical CCSN at the Galactic center (RA: \textit{$17^{h}45^{m}40.04^{s}$}, Dec: \textit{$-29^{\circ}00^{\prime}28.1^{\prime \prime}$}) of 10 kpc, which is shown as the triangle marker in Fig.~\ref{fig:pre_candidate}, the average number of $sn$IBD events for SN direction reconstruction is about 2000 in the NO case and 2400 in the IO case with SN neutrinos from the 13 $M_{\odot}$ Nakazato Model. As a result, the reconstructed direction uncertainty is about $26^\circ$ in the NO case and $23^{\circ}$ in the IO case as shown in the shadowed areas surrounded with the orange and green lines of Fig.~\ref{fig:pre_candidate}. Very close (e.g. closer than 1~kpc) SN would require special simulation tools and reconstruction methods to take into account the high event rates and possible pile-up effects. This situation will not be discussed in this paper.

 \section{Conclusion}
 \label{sec:con}
 
Neutrinos produced from a massive star before and during core collapse form a burst of the pre-SN and SN neutrinos, respectively. The early and prompt detection of the pre-SN and SN neutrinos provides a unique opportunity to realize the multi-messenger observation of the CCSN events. In this work, we describe the concept of the real-time monitoring system in JUNO and present its sensitivity.
The system consists of prompt monitors on the trigger boards and online monitors at the DAQ stage to ensure both alert speed and alert coverage of progenitor stars. A triggerless data processing scheme is also developed to limit the loss of CCSN-related information.

JUNO is unique and has good capability for the CCSN neutrino observation because of its large target mass, low background and multiple-flavor detection channels. In this work, the bench marking background rates and the sensitivity of the prompt monitor and online monitor have been quantified with full-chain simulated event samples. Assuming a FAR of 1/year, the online pre-SN monitor is sensitive to pre-SN stars up to 1.6~kpc (0.9~kpc) for a star of 30~$M_{\odot}$ in the NO (IO) case. 
For a Betelgeuse-like star at 0.15~kpc with 15 $M_{\odot}$ (0.25~kpc with 30 $M_{\odot}$), a pre-SN alert could be released 141 (73) hr or 84 (2.6) hr before the core collapse in the NO (IO) case. Both the prompt monitor and online monitor provide comparable sensitivity to the SN neutrinos, where a CCSN up to 370 kpc (360 kpc) could be covered for the NO (IO) case. The intrinsic alert latency is expected to be short in the prompt monitor for SN neutrinos, which is about ~20 ms for a 13 $M_{\odot}$ progenitor star with Nakazato model at 10~kpc.
We have also evaluated the CCSN pointing ability by using the accumulated IBD events from pre-SN or SN neutrinos. For a Betelgeuse-like star, the pointing ability with the pre-SN events is about $56^{\circ}$ ($81^{\circ}$) in the NO (IO) case for the 15 $M_{\odot}$ Patton model.
For a typical CCSN at 10 kpc, the pointing ability with SN neutrinos is about $26^\circ$ ($23^{\circ}$) in the NO (IO) case using the 13 $M_{\odot}$ Nakazato model.

In the future, better performance of the real-time system could be achieved by developing more complex monitoring algorithms and by optimizing the candidate selection with all possible interaction channels.
In addition to the prompt and online monitors based on the global trigger system, there will be an independent Multi-messenger trigger system, with the goal of achieving an ultra-low detection threshold for events of energies O(10) keV. 
The primary goal of the Multi-messenger trigger system is to allow for the detection of low-energy transient neutrino signals, and provide the CCSN monitor with an extended energy band of all-flavor neutrinos to maximize the alert ability to the global network of optical, gravitational and neutrino telescopes.
Therefore, these monitoring systems are effective in early alerts and the information of the CCSN directionality would play an important role in the multi-messenger observation of the next CCSN.

\section*{Acknowledgements}

We are grateful for the ongoing cooperation from the China General Nuclear Power Group.
This work was supported by
the Chinese Academy of Sciences,
the National Key R\&D Program of China,
the CAS Center for Excellence in Particle Physics,
Wuyi University,
and the Tsung-Dao Lee Institute of Shanghai Jiao Tong University in China,
the Institut National de Physique Nucl\'eaire et de Physique de Particules (IN2P3) in France,
the Istituto Nazionale di Fisica Nucleare (INFN) in Italy,
the Italian-Chinese collaborative research program MAECI-NSFC,
the Fond de la Recherche Scientifique (F.R.S-FNRS) and FWO under the ``Excellence of Science – EOS” in Belgium,
the Conselho Nacional de Desenvolvimento Cient\'ifico e Tecnol\`ogico in Brazil,
the Agencia Nacional de Investigacion y Desarrollo and ANID - Millennium Science Initiative Program - ICN2019\_044 in Chile,
the Charles University Research Centre and the Ministry of Education, Youth, and Sports in Czech Republic,
the Deutsche Forschungsgemeinschaft (DFG), the Helmholtz Association, and the Cluster of Excellence PRISMA+ in Germany,
the Joint Institute of Nuclear Research (JINR) and Lomonosov Moscow State University in Russia,
the joint Russian Science Foundation (RSF) and National Natural Science Foundation of China (NSFC) research program,
the MOST and MOE in Taiwan,
the Chulalongkorn University and Suranaree University of Technology in Thailand,
University of California at Irvine and the National Science Foundation in USA.

\end{document}